\documentclass[twocolumn,showpacs,floatfix,eqsecnum,prb]{revtex4}

\usepackage{amssymb}
\usepackage{amsmath}
\usepackage[dvips]{epsfig}\usepackage{float}
\usepackage{latexsym}\usepackage{revsymb}\usepackage{color}
\usepackage{multirow}\usepackage{array}

\usepackage{float}

\begin{document}

\title{Topological Defects and Gapless Modes in Insulators and Superconductors}
\author{Jeffrey C.Y. Teo and C.L. Kane}
\affiliation{Dept. of Physics and Astronomy, University of
Pennsylvania, Philadelphia, PA 19104}

\begin{abstract}
We develop a unified framework to classify topological defects in
insulators and superconductors described by spatially modulated Bloch
and Bogoliubov de Gennes Hamiltonians.  We consider Hamiltonians
$\mathcal{H}({\bf k},{\bf r})$ that vary slowly with adiabatic
parameters ${\bf r}$ surrounding the defect and belong to any of the ten
symmetry classes defined by time reversal symmetry and particle-hole symmetry.
The topological classes for such defects are identified, and explicit formulas
for the topological invariants are presented.
We introduce a generalization of the bulk-boundary correspondence that relates
the topological classes to defect Hamiltonians to the presence of protected
gapless modes at the defect.  Many examples of line and point defects in
three dimensional systems will be discussed.  These can host one dimensional chiral
Dirac fermions, helical Dirac fermions, chiral Majorana fermions and
helical Majorana fermions, as well as zero dimensional chiral and Majorana zero modes.  This approach
can also be used to classify temporal pumping cycles, such as the Thouless
charge pump, as well as a fermion parity pump, which is related to
the Ising non-Abelian statistics of defects that support Majorana zero modes.
\end{abstract}

\pacs{73.20.-r,73.43.-f, 71.10.Pm, 74.45.+c}
\maketitle

\section{Introduction}
\label{sec:introduction}

The classification of electronic phases according to topological
invariants is a powerful tool for understanding and predicting the
behavior of matter.  This approach was pioneered by Thouless, et
al.\cite{tknn}(TKNN), who identified the integer topological invariant characterizing
the two dimensional (2D) integer quantum Hall state.
The TKNN invariant $n$ gives the Hall conductivity $\sigma_{xy}=ne^2/h$
and characterizes the Bloch Hamiltonian ${\cal H}({\bf k})$, defined
as a function of ${\bf k}$ in the magnetic Brillouin zone.  It may
be expressed as the first Chern number associated with the
Bloch wavefunctions of the occupied states.
A fundamental consequence of this topological classification is the
{\it bulk-boundary correspondence}, which relates the topological
class of the bulk system to the number of gapless chiral fermion edge
states on the sample boundary.

Recent interest in topological states\cite{qizhang10,moore10,hk10} has been
stimulated by the realization that the combination of time
reversal symmetry and the spin orbit interaction
can lead to topological insulating electronic phases
\cite{km05a,km05b,moorebalents07,roy1,fkm07,qihugheszhang08} and by the
prediction\cite{bhz06,fukane07,zhang09} and observation
\cite{konig1,konig2,roth,hsieh08,hsieh09a,roushan09,xia09a,hor09,chen09,hsieh09b,park10,
alpichshev10,hsieh09c}
of these phases in real materials.
A topological insulator is a two or three dimensional
material with a bulk energy gap that has gapless
modes on the edge or surface that are protected by time reversal symmetry.
The bulk boundary correspondence relates these modes to a
$\mathbb{Z}_2$ topological invariant characterizing time reversal
invariant Bloch Hamiltonians.
Signatures of these protected boundary modes have been observed in transport
experiments on 2D HgCdTe quantum wells\cite{konig1,konig2,roth} and in photoemission and
STM experiments
on 3D crystals of Bi$_{1-x}$Sb$_x$\cite{hsieh08,hsieh09a,roushan09},
Bi$_2$Se$_3$\cite{xia09a}, Bi$_2$Te$_3$\cite{chen09,hsieh09b,alpichshev10} and
Sb$_2$Te$_3$\cite{hsieh09c}.
Topological insulator behavior has also been predicted in other classes of
materials with strong spin orbit interactions
\cite{shitade09,pesin10,chadov10,lin10a,lin10b,lin10c,yan10}.

Superconductors, described within a Bogoliubov de Gennes (BdG) framework can
similarly be classified topologically\cite{roy08,schnyder08,kitaev09,qi09}.
The Bloch-BdG Hamiltonian
${\cal H}_{BdG}({\bf k})$ has a structure similar to an ordinary
Bloch Hamiltonian, except that it has an exact particle-hole symmetry
that reflects the particle-hole redundancy inherent to the BdG
theory.  Topological superconductors are
also characterized by gapless boundary modes.  However, due to the
particle-hole redundancy, the boundary excitations
are Majorana fermions.  The simplest model topological superconductor is a
weakly paired spinless
$p$ wave superconductor in 1D\cite{kitaev00}, which has zero energy Majorana bound
states at its ends.  In 2D, a weakly paired $p_x+ip_y$ superconductor
has a chiral Majorana edge state\cite{readgreen}.  Sr$_2$RuO$_4$ is believed to exhibit a
triplet $p_x+ip_y$ state\cite{mackenzie03}.  The spin degeneracy, however, leads to a doubling
of the Majorana edge states.
Though undoubled topological superconductors remain to be
discovered experimentally, superfluid $^3$He B is a related
topological phase\cite{volovik03,roy08,schnyder08,qi09,volovik09}
and is predicted to exhibit 2D gapless Majorana modes on its
surface.  Related ideas have also been used to topologically classify
Fermi surfaces\cite{horava}.

Topological insulators and superconductors fit together into an
elegant mathematical framework that generalizes the above
classifications\cite{schnyder08,kitaev09}.  The topological classification of a general Bloch
or BdG theory is specified by the dimension $d$ and the 10 Altland
Zirnbauer symmetry classes\cite{altland97} characterizing the presence or absence of
particle-hole, time reversal and/or chiral symmetry.  The topological
classifications, given by $\mathbb{Z}$, $\mathbb{Z}_2$ or $0$ show a regular
pattern as a function of symmetry class and $d$, and can be arranged
into a {\it periodic table} of topological insulators and
superconductors.  Each non trivial entry in the table is predicted,
via the bulk-boundary correspondence, to have gapless boundary
states.

Topologically protected zero modes and gapless states can also occur at
topological defects, and have deep implications in both field theory and
condensed matter physics\cite{jackiwrebbi,jackiwrossi,ssh,volovik03}.
A simple example is the zero energy Majorana mode
that occurs at a vortex in a $p_x+ip_y$ superconductor\cite{readgreen}.
Similar Majorana bound states can
be engineered using three dimensional heterostructures that combine ordinary
superconductors and topological insulators\cite{fukane08}, as well as semiconductor
structures that combine superconductivity, magnetism and strong
spin orbit interactions\cite{sau10,alicea10,lutchyn10,oreg10}.   Recently, we
showed that the existence of a Majorana bound state at a point defect in a
three dimensional Bogoliubov de Gennes theory is related to a
$\mathbb{Z}_2$ topological invariant that characterizes a family of
Bogoliubov de Gennes Hamiltonians ${\cal H}_{BdG}({\bf k},{\bf r})$
defined for $\bf r$ on a surface
surrounding the defect\cite{teokane10}.  This suggests that a more general formulation
of topological defects and their corresponding gapless modes should
be possible.

\begin{figure}
\epsfxsize=3.2in
\epsfbox{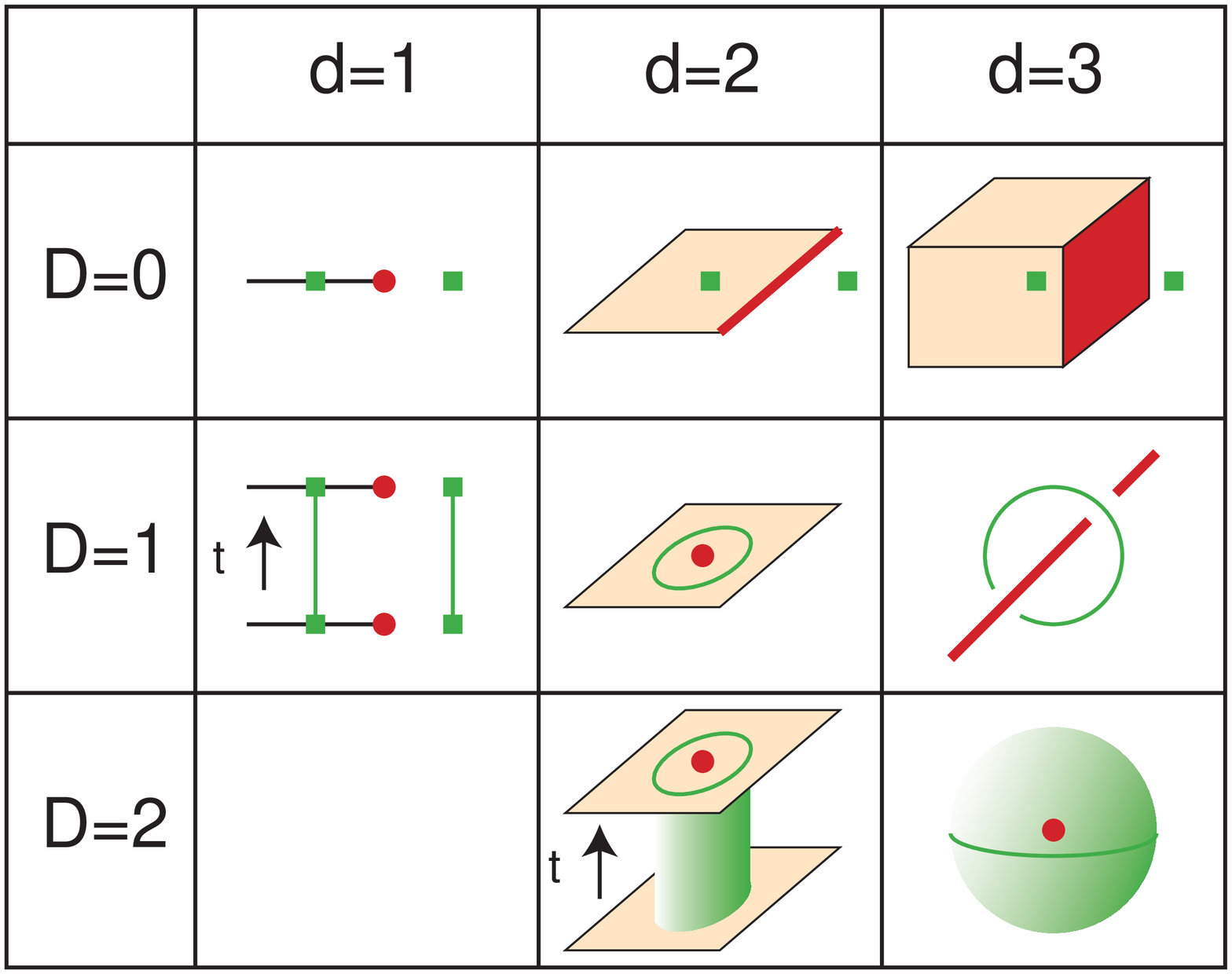}
\caption{Topological defects characterized by a $D$ parameter family of $d$ dimensional
Bloch-BdG Hamiltonians.  Line defects correspond to $d-D=2$,
while point defects correspond to $d-D=1$.  Temporal cycles for point defects correspond
to $d-D=0$.}
\label{figure1}
\end{figure}

In this paper we develop a general theory of topological defects and their
associated gapless modes in Bloch and Bloch-BdG theories in all symmetry classes.
As in Ref. \onlinecite{teokane10}, we assume that far away
from the defect the Hamiltonian varies slowly in real space, allowing us to
consider adiabatic changes of the Hamiltonian as a function of the real space position
${\bf r}$.  We thus
seek to classify Hamiltonians ${\cal H}({\bf k},{\bf r})$, where ${\bf
k}$ is defined in a $d$ dimensional Brillouin zone (a torus $T^d$),
and ${\bf r}$ is defined on a $D$ dimensional surface $S^D$
surrounding the defect.
A similar approach can be used to classify cyclic
temporal variations in the Hamiltonian, which define adiabatic pumping cycles.
Hereafter we will drop the BdG subscript on the Hamiltonian with the understanding
that the symmetry class dictates whether it is a Bloch or BdG Hamiltonian.

In Fig. \ref{figure1} we illustrate the types of topological defects that can occur in $d=1$,
$2$ or $3$.  For $D=0$ we regard $S^0$ as two points ($\{-1,+1\}$).  Our
topological classification then classifies the {\it difference} of ${\cal H}({\bf
k},+1)$ and ${\cal H}({\bf k},-1)$.  A non trivial difference corresponds to an
interface between two topologically distinct phases.
For $D=1$ the one parameter families of Hamiltonians
describe line defects in $d=3$ and point defects in $d=2$.  For $d=1$ it could
correspond to an adiabatic temporal cycle $H({\bf k},t)$.  Similarly for $D=2$,
the two parameter family describes a point defect for $d=3$ or an adiabatic
cycle for a point defects in $d=2$.

Classifying the $D$ parameter families of $d$ dimensional Bloch-BdG Hamiltonians
subject to symmetries leads to a
generalization of the periodic table discussed above.  The original table
corresponds to $D=0$.  For $D>0$ we find that for a given symmetry class the
topological classification ($\mathbb{Z}$, $\mathbb{Z}_2$ or $0$) depends only on
\begin{equation}
\delta = d - D.
\label{delta}
\end{equation}
Thus, all line defects with $\delta = 2$ have the same topological
classification, irrespective of $d$, as do  point defects with $\delta = 1$ and
pumping cycles with $\delta = 0$.
Though the classifications depend only on $\delta$, the {\it formulas} for the
topological invariants depend on both $d$ and $D$.

This topological classification of ${\cal H}({\bf k},{\bf r})$
suggests a generalization of the
bulk-boundary correspondence that relates the topological class of
the Hamiltonian characterizing the defect to the structure of the
protected modes associated with the defect.  This has a
structure reminiscent of a mathematical {\it index theorem}\cite{nakahara} that
relates a topological index to an analytical index that counts the
number of zero modes\cite{jackiwrebbi,jackiwrossi,volovik03,goldstone,weinberg,witten,witten82,davis}.
In this paper we will not attempt to {\it prove} the
index theorem.  Rather, we will observe that the topological classes
for ${\cal H}({\bf k},{\bf r})$ coincide with the expected classes of
gapless defect modes.  In this regards the dependence of the
classification on $\delta$ in \eqref{delta} is to be expected.
For example, a point defect at the end of a one
dimensional system ($\delta = 1-0$) has the same classification as a point
defects in two dimensions ($\delta = 2-1$) and three dimensions ($\delta =
3-2$).

We will begin in section \ref{sec:periodictable} by describing the generalized periodic table.  We
will start with a review of the Altland Zirnbauer symmetry classes\cite{altland97} and a
summary of the properties of the table.  In Appendix \ref{appendix:periodicities} we will justify
this generalization of the table by introducing
a set of mathematical mappings that relate Hamiltonians in different dimensions and different
symmetry classes.  In addition to establishing that the classifications depend
only on $\delta=d-D$, these mappings allow other features of the table,
already present for $D=0$ to be easily understood, such as the pattern in which the
classifications vary as a function of symmetry class as well
as the Bott periodicity of the classes as a function of $d$.

In section \ref{sec:linedefects} and \ref{sec:pointdefect} we will outline the physical consequences of this theory by
discussing a number of examples of line and point defects in different symmetry
classes and dimensions.  The simplest example is that of a line defect in a
3D system with no symmetries.  In section \ref{sec:linechiraldirac} we will show that the presence
of a 1D {\it chiral Dirac
fermion} mode (analogous to an integer quantum Hall edge state) on the defect
is associated with an integer topological invariant that may be interpreted
as the winding number of the ``$\theta$" term that characterizes the
magnetoelectric polarizability\cite{qihugheszhang08}.  This description unifies a number of methods
for ``engineering" chiral Dirac fermions, which will be described in several
illustrative examples.

Related topological invariants and illustrative
examples will be presented in Sections \ref{sec:linechiralmajorana}-\ref{sec:lineclassc}
for line defects in other
symmetry classes that are associated with gapless 1D helical Dirac
fermions, 1D chiral Majorana fermions and 1D helical Majorana fermions.
In section \ref{sec:pointdefect} we will consider point defects in 1D models with chiral symmetry
such as the Jackiw Rebbi model\cite{jackiwrebbi} or the Su, Schrieffer, Heeger model\cite{ssh},
and in superconductors without chiral symmetry that exhibit Majorana
bound states or Majorana doublets.   These will also be related to the early
work of Jackiw and Rossi\cite{jackiwrossi} on Majorana modes at point defects in a model with chiral
symmetry.

Finally, in Section \ref{sec:pump} we will regard ${\bf r}$ as including a temporal variable,
and apply the considerations in this paper to
classify cyclic pumping processes.  The Thouless charge pump\cite{thouless,thoulessniu}
corresponds to a non trivial cycle in a
system with no symmetries and $\delta=0$ ($d=D=1$).  A similar pumping scenario
can be applied to superconductors and defines a {\it fermion parity pump}.
This, in turn, is related to the non-Abelian statistics of Ising anyons, and
provides a framework for understanding braidless operations on systems of three
dimensional superconductors hosting Majorana fermion bound states.
Details of several technical calculations can be found in the
Appendices.

An interesting recent preprint by Freedman et al.\cite{freedman10}, which appeared when
this manuscript was in its
final stages discusses some aspects of the classification of topological
defects in connection with a rigorous theory of non-Abelian statistics in higher
dimensions.

\section{Periodic Table for defect classification}
\label{sec:periodictable}

Table \ref{tab:periodic} shows the generalized periodic table for the classification
of topological defects in insulators and superconductors.  It
describes the equivalence classes of Hamiltonians ${\cal H}({\bf k},{\bf r})$,
that can be continuously deformed into one another without closing the energy
gap, subject to constraints of particle-hole and/or time reversal symmetry.
These are mappings from a {\it base space} defined by $({\bf
k},{\bf r})$ to a  {\it classifying space}, which characterizes
the set of gapped Hamiltonians.
In order to explain the table, we need to describe (i) the
symmetry classes, (ii) the base space, (iii) the
classifying space and (iv) the notion of stable equivalence.
The repeating patterns in the table will be discussed in Section \ref{sec:properties}.
Much of this section is a review of material in Refs. \onlinecite{schnyder08,kitaev09}.
What is new is the extension to $D>0$.

\begin{table}
\centering
\begin{ruledtabular}
\begin{tabular}{cc|ccc|cccccccc}
\multicolumn{5}{c|}{Symmetry } & \multicolumn{8}{c}{ $\delta = d - D$} \\
$s$& \multicolumn{1}{c}{AZ} &$\Theta^2$ &$\Xi^2$ & $ \Pi^2$ &
$0$ & $1$   &  $2$ &  $3$ &  $4$ &  $5$ & $6$ & $7$\\
 \hline
0 & A & $0$ & $0$ & $0$ &
   $\mathbb{Z}$ & $0$ & $\mathbb{Z}$ & $0$ & $\mathbb{Z}$ & $0$ & $\mathbb{Z}$ & $0$\\
1 & AIII & $0$ & $0$ & $1$ &
   $0$ & $\mathbb{Z}$ & $0$ & $\mathbb{Z}$ & $0$ & $\mathbb{Z}$ & $0$ & $\mathbb{Z}$\\
\hline
0 & AI &  $1$ & $0$ & $0$ &
     $\mathbb{Z}$ & $0$ & $0$ & $0$ & $2\mathbb{Z}$ & $0$ & $\mathbb{Z}_2$ & $\mathbb{Z}_2$\\
1 & BDI & $1$ & $1$ &$1$ &
     $\mathbb{Z}_2$ & $\mathbb{Z}$ & $0$ & $0$ & $0$ & $2\mathbb{Z}$ & $0$ & $\mathbb{Z}_2$\\
2 & D & $0$ & $1$ & $0$ &
     $\mathbb{Z}_2$ & $\mathbb{Z}_2$ & $\mathbb{Z}$ & $0$ & $0$ & $0$ & $2\mathbb{Z}$ & $0$\\
3 & DIII & $-1$ & $1$ & $1$ &
     $0$ & $\mathbb{Z}_2$ & $\mathbb{Z}_2$ & $\mathbb{Z}$ & $0$ & $0$ & $0$& $2\mathbb{Z}$ \\
4 & AII & $-1$ & $0$ & $0$ &
     $2\mathbb{Z}$ & $0$ & $\mathbb{Z}_2$ & $\mathbb{Z}_2$ & $\mathbb{Z}$ & $0$ & $0$ & $0$\\
5 & CII & $-1$ & $-1$ & $1$ &
     $0$ & $2\mathbb{Z}$ & $0$ & $\mathbb{Z}_2$ & $\mathbb{Z}_2$ & $\mathbb{Z}$ & $0$ & $0$\\
6 & C &  $0$ & $-1$ & $0$ &
     $0$ & $0$ & $2\mathbb{Z}$ & $0$ & $\mathbb{Z}_2$ & $\mathbb{Z}_2$ & $\mathbb{Z}$ & $0$\\
7 & CI &  $1$ & $-1$ & $1$ &
     $0$ & $0$ & $0$ & $2\mathbb{Z}$ & $0$ & $\mathbb{Z}_2$ & $\mathbb{Z}_2$ & $\mathbb{Z}$\\
\end{tabular}
\end{ruledtabular}
\caption{Periodic table for the classification of topological defects in insulators and
superconductors.  The rows correspond to the different Altland Zirnbauer (AZ) symmetry
classes, while the columns distinguish different dimensionalities, which depend
only on $\delta = d-D$.
 }
\label{tab:periodic}
\end{table}

\subsection{Symmetry Classes}
\label{sec:symmetryclasses}

The presence or absence of time reversal symmetry, particle-hole symmetry
and/or chiral symmetry define the 10 Altland Zirnbauer symmetry
classes\cite{altland97}.   Time reversal symmetry implies that
\begin{equation}
{\cal H}({\bf k},{\bf r}) = \Theta {\cal H}(-{\bf k},{\bf r})
\Theta^{-1},
\label{trsym}
\end{equation}
where the anti unitary time reversal operator may be written
$\Theta = e^{i\pi S^y/\hbar}K$.   $S^y$ is the spin and
$K$ is complex conjugation.
For spin 1/2 fermions, $\Theta^2 =
-1$, which leads to Kramers theorem.
In the absence of a spin orbit interaction, the extra invariance of the Hamiltonian
under rotations in spin space allows an additional time reversal operator
$\Theta' = K$ to be defined, which satisfies $\Theta'^2 = +1$.

Particle-hole symmetry is expressed by
\begin{equation}
{\cal H}({\bf k},{\bf r}) = -\Xi {\cal H}(-{\bf k},{\bf r})
\Xi^{-1},
\label{phsym}
\end{equation}
where $\Xi$ is the anti unitary particle-hole operator.  Fundamentally,
$\Xi^2 = +1$.  However, as was the case for $\Theta$, the absence of spin orbit
interactions introduces an additional particle hole symmetry, which can satisfy
$\Xi^2=-1$.

Finally, chiral symmetry is expressed by a unitary operator $\Pi$,
satisfying
\begin{equation}
{\cal H}({\bf k},{\bf r}) = -\Pi {\cal H}({\bf k},{\bf r})
\Pi^{-1}.
\label{chiralsym}
\end{equation}
A theory with both particle-hole
and time reversal symmetry
automatically has a chiral symmetry $\Pi = e^{i\chi} \Theta \Xi$.  The phase
$\chi$ can be chosen so that $\Pi^2=1$.

Specifying $\Theta^2 = 0, \pm 1$, $\Xi^2 = 0, \pm 1$ and
$\Pi^2 = 0, 1$ (here $0$ denotes the absence of symmetry)
defines the 10 Altland Zirnbauer symmetry classes.
They can be divided into two groups:  8 {\it real} classes that
have anti unitary symmetries $\Theta$ and or $\Xi$ plus 2 {\it
complex} classes that do not have anti unitary symmetries.
Altland and Zirnbauer's notation for these classes, which is based
on Cartan's classification of symmetric spaces, is shown in the left
hand part of Table \ref{tab:periodic}.

\begin{figure}
\epsfxsize=2.0in
\epsfbox{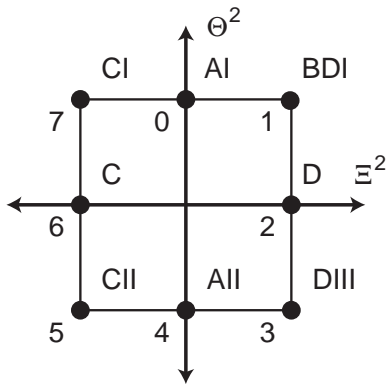}
\caption{The 8 Real symmetry classes that involve
the anti unitary symmetries $\Theta$ (time reversal) and/or
$\Xi$ (particle-hole) are specified by the values
of $\Theta^2 = \pm 1$ and $\Xi^2=\pm 1$.  They can be visualized on an eight
hour ``clock". }
\label{clock}
\end{figure}

To appreciate the mathematical structure of the 8 real symmetry classes it is
helpful to picture them on an 8 hour ``clock", as shown in Fig. \ref{clock}.
The $x$ and $y$ axes of the clock represent the values of $\Xi^2$
and $\Theta^2$.  The ``time" on the clock can be represented by an
integer $s$ defined modulo 8.   Kitaev\cite{kitaev09} used a slightly different
notation to label the symmetry classes.  In his formulation, class D
is described by a real Clifford algebra with no constraints, and in the
other classes Clifford algebra elements are constrained to
anticommute with $q$ positive generators.  The two formulations are
related by $s = q+2$ mod $8$.
The complex symmetry classes can similarly be indexed by an integer
$s$ defined modulo 2.  For all classes, the presence of chiral
symmetry is associated with odd $s$.

\subsection{Base space, Classifying space and stable equivalence}
\label{sec:basespace}

The Hamiltonian is defined on a base space composed of momentum
${\bf k}$, defined in a $d$ dimensional Brillouin zone $T^d$
 and real space degrees of freedom ${\bf r}$ in a sphere $S^D$ (or
 $S^{D-1}\times S^1$ for an adiabatic cycle).  The total base space is therefore
$T^d \times S^D$ (or $T^d \times S^{D-1} \times S^1$).  As in Ref. \onlinecite{kitaev09}, we will
simplify the topological classification by treating the base space as a sphere $S^{d+D}$.
The ``strong" topological invariants that characterize the sphere will also characterize
$T^d \times S^D$.  However, there may be additional topological structure
in $T^d \times S^D$ that is absent in $S^{d+D}$.  These
correspond to ``weak" topological invariants.  For $D=0$ these arise in layered
structures.  A weak topological insulator, for example, can be understood as
a layered two dimensional topological insulator.  There are similar layered
quantum Hall states.  For $D\ne 0$, then there will also be a ``weak" invariant
if the Hamiltonian ${\cal H}({\bf k},{\bf r}_0)$ for fixed ${\bf r}={\bf r}_0$
is topologically non trivial.
As is the case for the
classification of bulk phases $D=0$, we expect that the topologically protected
gapless defect modes are associated with the strong topological invariants.

The set of Hamiltonians that preserve the energy gap
separating positive and negative energy states can be simplified
without losing any topological information.  Consider the retraction
of the original Hamiltonian ${\cal H}({\bf k},{\bf r})$ to a simpler
Hamiltonian whose eigenvalue spectrum is ``flattened", so that the
positive and negative energy states all have the same energy $\pm E_0$.
The flattened Hamiltonian is then specified the set of all $n$ eigenvectors (defining a $U(n)$
matrix) modulo unitary rotations within the $k$ conduction bands or the $n-k$ valence
bands.  The flattened Hamiltonian can thus be identified with a point in the Grassmanian
manifold,
\begin{equation}
G_{n,k} = U(n)/U(k)\times U(n-k).
\label{grassmanian}
\end{equation}

It is useful to broaden the notion of topological equivalence
to allow for the presence of extra trivial energy bands.  Two families of Hamiltonians
are {\it stably equivalent} if they can be deformed into one another after adding an arbitrary
number of trivial bands.  Thus, trivial insulators with
different numbers of core energy levels are stably equivalent.
Stable equivalence can be implemented by considering an expanded {\it classifying
space} that includes an infinite number of extra conduction and valence bands,
${\cal C}_0 = U / U \times U \equiv  \bigcup_{k=0}^\infty
G_{\infty,k}$.

With this notion of stable equivalence, the equivalence
classes of Hamiltonians ${\cal H}({\bf k},{\bf r})$ can be formally
added and subtracted.  The addition of two classes, denoted $[{\cal H}_1] + [{\cal
H}_2]$ is formed by simply combining two independent Hamiltonians
into a single Hamiltonian given by the matrix direct sum, $[{\cal H}_1
\oplus {\cal H}_2]$.   Additive inverses are constructed
through reversing conduction and valence bands,
$[{\cal H}_1]-[{\cal H}_2] = [{\cal H}_1 \oplus -{\cal H}_2]$.
$[{\cal H} \oplus -{\cal H}]$ is guaranteed to the be trivial class $[0]$.
Because of this property, the stable equivalence classes form
an Abelian group, which is the key element of K theory\cite{karoubi,lawson,atiyah94}.

Symmetries impose constraints on the classifying space.
For the symmetry classes with chiral symmetry,
\eqref{chiralsym} restricts $n=2k$ and the
classifying space to a subset ${\cal C}_1 = U(\infty) \subset U/U \times U$.
The anti unitary symmetries (\ref{trsym},\ref{phsym}) impose further constraints.
At the special points where ${\bf k}$ and $-{\bf k}$ coincide, the
allowed Hamiltonians are described by the 8 classifying spaces ${\cal R}_q$ of Real K
theory.

\subsection{Properties of the periodic table}
\label{sec:properties}

For a given symmetry class $s$, the topological classification of defects is given by the set of
stable equivalence classes of maps from the base space $({\bf k},{\bf r}) \in S^{D+d}$
to the classifying space, subject to the symmetry constraints.  These form the K group, which we denote
as $K_{\mathbb{C}}(s;D,d)$ for the complex symmetry classes and $K_{\mathbb{R}}(s;D,d)$ for the real
symmetry classes.   These are listed in Table \ref{tab:periodic}.

Table \ref{tab:periodic} exhibits many remarkable patterns.  Many can be understood
from the following basic periodicities,
\begin{eqnarray}
K_{\mathbb{F}}(s;D,d + 1) &=& K_{\mathbb{F}}(s- 1;D,d) \label{period1},\\
K_{\mathbb{F}}(s;D + 1, d) &=& K_{\mathbb{F}}(s+ 1;D,d)\label{period2}.
\end{eqnarray}
Here $s$ is understood to be defined modulo 2 for $\mathbb{F} = \mathbb{C}$
and modulo 8 for  $\mathbb{F} = \mathbb{R}$.  We will establish these
identities mathematically in Appendix \ref{appendix:periodicities}.  The basic idea is to start with some
Hamiltonian in some symmetry class $s$ and dimensionalities $D$ and $d$.  It is
then possible to explicitly construct two new Hamiltonians in one higher
dimension which have either (i) $d \rightarrow d+1$ or (ii) $D \rightarrow D+1$.  These new
Hamiltonians belongs to new symmetry classes that are shifted by one ``hour" on the
symmetry clock and characterized by (i) $s \rightarrow
s+1$ or (ii) $s \rightarrow s-1$.  We then go on to show that this construction
defines a 1-1 correspondence between the equivalence classes of Hamiltonians
with the new and old symmetry classes and dimensions, thereby establishing
\eqref{period1} and \eqref{period2}.

The periodicities
\eqref{period1} and \eqref{period2} have a number of consequences.  The most important for our
present purposes is they can be combined to give
\begin{equation}
K_\mathbb{F}(s;D+1,d+1) = K_\mathbb{F}(s;D,d).
\end{equation}
This $(1,1)$ periodicity shows that the
dependence on the dimensions $d$ and $D$ only occurs via $\delta =
d-D$.  Thus the dependence of the classifications on $D$ can be
deduced from the table for $D=0$.  This is one of our central
results.

In addition, the periodicities (\ref{period1},\ref{period2}) explain
other features of the table that are already present for $D=0$.  In
particular, the fact that $s$ is defined modulo 2 (8) for the complex
(real) classes leads directly to the Bott periodicity of the
dependence of the classifications on $d$:
\begin{eqnarray}
K_{\mathbb{C}}(s;D,d+2) &=& K_{\mathbb{C}}(s;D,d), \\
K_{\mathbb{R}}(s;D,d+8) &=& K_{\mathbb{R}}(s;D,d).
\end{eqnarray}
Moreover, (\ref{period1},\ref{period2}) shows that $K_a(s;D,d)$ depends only
on $d-D-s$.  This explains the diagonal pattern
in Table \ref{tab:periodic}, in which the dependence of the classification on $d$ is
repeated in successive symmetry classes.  Thus, the entire table
could be deduced from a single row.

Equations \eqref{period1} and \eqref{period2} do not explain the
pattern of classifications within a single row.  Since this is a well
studied math problem there are many routes to the answer\cite{bott,JMilnor,lawson}.  One
approach is to notice that for $d=0$, $K_\mathbb{F}(s,D,0)$ is simply the
$D$'th homotopy group of the appropriate classifying space which
incorporates the symmetry constraints.  For example, for class BDI
($s=1$, $\Xi^2 = +1$, $\Theta^2=+1$) the classifying space is the orthogonal group
$O(\infty)$.  Then, $K_{\mathbb{R}}(1,D,0) = \pi_D(O(\infty))$,
which are well known.   This implies
\begin{equation}
K_{\mathbb{R}}(s;D,d) =\pi_{s+D-d-1}(O(\infty)).
\end{equation}

Additional insight can be obtained by
examining the interconnections between different elements of the
table.  For example, the
structure within a column can be analyzed by considering the effect
of ``forgetting" symmetries.  Hamiltonians belonging to the real
chiral (non chiral) classes are automatically in complex class AIII (A).
There are therefore K group homomorphisms that send any real entries in table \ref{tab:periodic}
to complex ones directly above.  In particular, as detailed in Appendix \ref{appendix:representatives}
 this distinguishes the
$\mathbb{Z}$ and $2\mathbb{Z}$ entries, which indicate the possible values of
Chern numbers (or $U(n)$ winding numbers) for even (or odd) $\delta$.
In addition, the dimensional reduction arguments given in
Refs. \onlinecite{qihugheszhang08,ryu10} lead to a dimensional hierarchy, which helps to explain the
pattern within a single row as a function of $d$.

\section{Line defects}
\label{sec:linedefects}

Line defects can occur at the edge of a 2D system ($\delta = 2-0$) or
in a 3D system ($\delta = 3-1$).  From Table \ref{tab:periodic}, it can be seen that there are
five symmetry classes which can host non trivial line defects.  These are
expected to be associated with gapless fermion modes bound to the defect.
Table \ref{linetab} lists non trivial classes, along with the character of the
associated gapless modes.  In the following subsections we will discuss each of
these cases, along with physical examples.

\begin{table}
\centering
\begin{ruledtabular}
\begin{tabular}{ccl}
Symmetry  & Topological classes & 1D Gapless Fermion modes\\
\hline
A & $\mathbb{Z}$ &  Chiral Dirac \\
D & $\mathbb{Z}$ &  Chiral Majorana \\
DIII & $\mathbb{Z}_2$ &  Helical Majorana \\
AII & $\mathbb{Z}_2$ &  Helical Dirac  \\
C & $2\mathbb{Z}$ &   Chiral Dirac
\end{tabular}
\end{ruledtabular}
\caption{Symmetry classes that support topologically non trivial line defects and their associated
protected gapless modes. }
\label{linetab}
\end{table}

\subsection{Class A: Chiral Dirac Fermion}
\label{sec:linechiraldirac}

\subsubsection{Topological Invariant}
\label{sec:linechiraldiracinvariant}

A line defect in a generic 3D Bloch band theory with no symmetries is associated
with an integer topological invariant.  This determines the number of
chiral Dirac fermion modes associated with the defect.
Since ${\cal H}({\bf k},{\bf r})$ is defined on a
compact 4 dimensional space, this invariant is naturally expressed as a second
Chern number,
\begin{equation}
n = {1\over {8\pi^2}} \int_{T^3\times S^1} {\rm Tr}[{\cal F}\wedge{\cal
F}],
\label{2ndchern}
\end{equation}
where
\begin{equation}
{\cal F} = d{\cal A} + {\cal A} \wedge {\cal A}
\label{F(A)}
\end{equation}
is the curvature form
associated with the non-Abelian Berry's
connection ${\cal A}_{ij} = \langle u_i|d u_j\rangle$ characterizing the
valence band eigenstates $|u_j({\bf k},s)\rangle$ defined on the loop $S^1$ parameterized
by $s$.

It is instructive to rewrite this
as an integral over $s$ of a quantity associated with the local band structure.
To this end, it is useful to write ${\rm Tr}[{\cal F}\wedge{\cal F}] = d
{\cal Q}_3$, where the Chern Simons 3 form is,
\begin{equation}
{\cal Q}_3 = {\rm Tr}[{\cal A}\wedge d{\cal A} + {2\over 3} {\cal A}\wedge{\cal
A}\wedge{\cal A}].
\label{q3}
\end{equation}
Now divide the integration volume into thin slices, $T^3 \times \Delta
S^1$, where $\Delta S^1$ is the interval between $s$ and $s+\Delta s$.
In each slice, Stokes' theorem may be used to write the integral as a surface
integral over the surfaces of the slice at $s$ and $s+\Delta s$.  In this
manner, Eq. \ref{2ndchern} may be written
\begin{equation}
n = {1\over {2\pi}} \oint_{S^1} ds {d\over {ds}}\theta(s),
\label{n(theta)}
\end{equation}
where
\begin{equation}
\theta(s) = {1\over {4\pi}} \int_{T^3} {\cal Q}_3({\bf k},s).
\label{theta(s)}
\end{equation}
Eq. \ref{theta(s)} is precisely the Qi, Hughes, Zhang formula\cite{qihugheszhang08} for the
``$\theta$" term that characterizes the magnetoelectric response of a band
insulator.  $\theta = 0$ for an ordinary time reversal invariant insulator, and
$\theta = \pi$ in a strong topological insulator.  If parity and time reversal
symmetry are broken then $\theta$ can have any intermediate value.  We thus
conclude that the topological invariant associated with a line
defect, which determines the number of chiral fermion branches is given by the
winding number of $\theta$.

We now consider several examples of 3D line defects that are associated with
chiral Dirac fermions.

\subsubsection{Dislocation in a 3D Integer quantum Hall state}
\label{sec:qhedislocation}

\begin{figure}
\epsfxsize=2.5in
\epsfbox{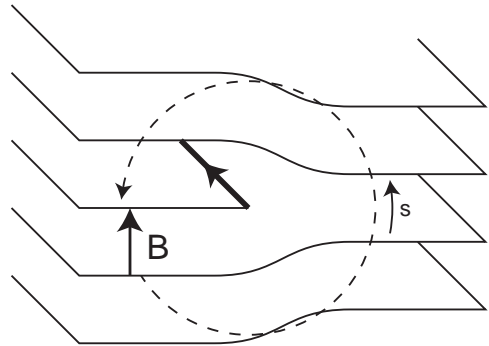}
\caption{A line dislocation in a three dimensional quantum Hall state characterized
by Burgers vector ${\bf B}$.}
\label{dislocation}
\end{figure}

A three dimensional integer quantum Hall state can be thought of as a layered
version of the two dimensional integer quantum Hall state.   This can be understood most
simply by considering the extreme limit where the layers are completely
decoupled 2D systems.   A line dislocation, as shown in Fig. \ref{dislocation}
will then involve an edge of one of the planes and be associated with a chiral
fermion edge state.  Clearly, the chiral fermion mode will remain when the layers
are coupled, provided the bulk gap remains finite.
Here we wish to show how the topological invariant
\eqref{2ndchern} reflects this fact.

On a loop surrounding the dislocation parameterized by $s\in[0,1]$
we may consider a family of Hamiltonians $H({\bf k},s)$ given by the Hamiltonian
of the original bulk crystal displaced by a distance $s{\bf B}$, where ${\bf
B}$ is a lattice vector equal to the Burgers vector of the defect.
The corresponding Bloch wavefunctions will thus be given by,
\begin{equation}
u_{m{\bf k},s}({\bf r}) = u^0_{m\bf k}({\bf r}- s {\bf B}),
\label{dislocationu}
\end{equation}
where $u^0_{m\bf k}({\bf r})$ are Bloch functions for the original crystal.
It then follows that the Berry's connection is
\begin{equation}
{\cal A} = {\cal A}^0 +  {\bf B}\cdot ({\bf k} - {\bf a}^p({\bf k})) ds,
\label{dislocationa}
\end{equation}
where
${\cal A}^0_{mn}({\bf k}) =
\langle u^0_{m\bf k}|\nabla_{\bf k} |u^0_{n\bf k}\rangle\cdot d{\bf k}$
 and
\begin{equation}
{\bf a}^p_{mn}({\bf k}) = \langle u^0_{m\bf k}| (\nabla_{\bf r} + {\bf k} ) |u^0_{n\bf
k}\rangle.
\end{equation}
With this definition, ${\bf a}^p({\bf k})$ is a periodic function: ${\bf
a}^p({\bf k}+{\bf G}) = {\bf a}^p({\bf k})$ for any reciprocal lattice vector
${\bf G}$\cite{blount}.

If the crystal is in a three dimensional quantum Hall state, then the non zero
first Chern number is an obstruction to finding the globally continuous gauge
necessary to evaluate \eqref{theta(s)}.  We therefore use
\eqref{2ndchern}, which can be evaluated by noting that
\begin{equation}
{\rm Tr}[{\cal F}\wedge{\cal F}] =
  {\rm Tr}\left[{\bf B}\cdot\left(2 {\cal F}^0 \wedge d{\bf k} -
d [{\cal F}^0,{\bf a}^p]\right) \wedge ds \right].
\label{dislocationff}
\end{equation}
Upon integrating ${\rm Tr}[{\cal F}\wedge{\cal F}]$ the total
derivative term vanishes due to the periodicity of ${\bf a}^p$.
Evaluating the integral is then straightforward.  The integral over $s$
trivially gives $1$.  We are then left with
\begin{equation}
n = {1\over {2\pi}} {\bf B} \cdot {\bf G}_c,
\end{equation}
where
\begin{equation}
{\bf G}_c = {1\over {2\pi}}\int_{T^3} d{\bf k} \wedge {\rm Tr}[{\cal F}^0].
\end{equation}
${\bf G}_c$ is a reciprocal lattice vector that corresponds to the triad of
Chern numbers that characterize a 3D system.  For instance, in a cubic system
${\bf G}_c = (2\pi/a)(n_x,n_y,n_z)$, where, for example $n_z = (2\pi)^{-1}\int
{\rm Tr}[{\cal F}^0_{xy}] dk_x\wedge dk_y$, for any value of $k_z$.

An equivalent formulation is to characterize the displaced crystal in terms of $\theta$.
Though \eqref{theta(s)} can not be used, \eqref{2ndchern} and \eqref{n(theta)} can be used to implicitly define
$\theta$ up to an arbitrary additive constant,
\begin{equation}
\theta(s)= s {\bf B}\cdot{\bf G}_c.
\end{equation}

\subsubsection{Topological insulator heterostructures}
\label{sec:linechiraldirachetero}

\begin{figure}
\epsfxsize=3in
\epsfbox{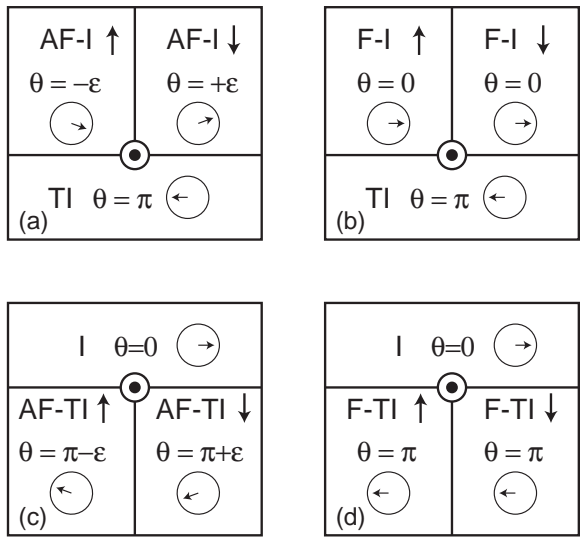}
\caption{Heterostructure geometries for chiral Dirac fermions.  (a) and (b) show
antiferromagnetic or ferromagnetic insulators on the surface of a topological insulator,
with chiral Dirac fermions at a domain wall.  (c) and (d) show a domain wall in an antiferromagnetic
or ferromagnetic topological insulator.  Chiral fermion modes are present when the domain wall intersects
the surface. }
\label{chiraldevice}
\end{figure}

Another method for engineering chiral Dirac fermions is use
heterostructures that combine topological insulators and magnetic materials.
The simplest version is a topological insulator coated with a magnetic film
that opens a time reversal symmetry breaking energy gap at the
surface.  A domain wall is then associated with a chiral fermion
mode.  In this section we will show how this structure, along with
some variants on the theme, fits into our general framework.  We first
describe the structures qualitatively, and then analyze a model that
describes them.

Fig. \ref{chiraldevice} shows four possible configurations.  Figs.
\ref{chiraldevice}(a,b) involve a topological insulator with magnetic materials
on the surface.  The magnetic material could be either
ferromagnetic or antiferromagnetic.  We distinguish these two cases based on whether
inversion symmetry is broken or not.   Ferromagnetism does not violate
inversion symmetry, while antiferromagnetism does (at least for inversion about
the middle of a bond).  This is relevant because  $\theta$ , discussed
above, is quantized unless {\it both} time reversal and inversion symmetries
are violated.  Of course, for a non centrosymmetric crystal inversion is
already broken, so the distinction is unnecessary.

Fig. \ref{chiraldevice}(a) shows a topological insulator capped with
antiferromagnetic insulators with $\theta = \pm \epsilon$
separated by a domain wall.  Around the junction where the
three regions meet $\theta$ cycles between $\pi$, $+\epsilon$ and $-\epsilon$.
Of course this interface structure falls outside the adiabatic regime that
\eqref{theta(s)} is based on.  However, it is natural to expect that the physics
would not change if the interface was ``smoothed out" with $\theta$ taking the
shortest smooth path connecting its values on either side of the interface.

Fig. \ref{chiraldevice}(b) shows a similar device with ferromagnetic insulators, for
which $\theta = 0$ or $\pi$.  In this case the adiabatic assumption again
breaks down, however, as emphasized in Ref. \onlinecite{qihugheszhang08},
the appropriate way to think
about the surface is that $\theta$ connects $0$ and $\pi$ along a path that is
determined by the sign of the induced gap, which in turn is related to the
magnetization.  In this sense, $\theta$ cycles by $2\pi$ around the junction.

In Figs. \ref{chiraldevice}(c,d) we consider topological insulators
which have a weak magnetic instability.  If in addition to time reversal,
inversion symmetry is broken, then
$\theta \sim \pi \pm \epsilon$.  Recently Li, Wang, Qi and Zhang\cite{li10} have considered
such materials in connection with a theory of a dynamical axion and
suggested that certain magnetically doped topological insulators may exhibit
this behavior.   They referred to such materials as topological magnetic insulators.
We prefer to call them {\it magnetic topological insulators} because as
magnetic insulators they are topologically trivial.  Rather, they are
topological insulators to which magnetism is added.  Irrespective of the name,
such materials would be extremely interesting to study, and as we discuss
below, may have important technological utility.

Fig. \ref{chiraldevice}(c) shows two antiferromagnetic topological insulators
with $\theta = \pi \pm \epsilon$ separated by a domain wall, and Fig. \ref{chiraldevice}(d)
shows a similar device with ferromagnetic topological insulators.  They form an
interface with an insulator, which could be vacuum.  Under the same continuity
assumptions as above the junction where the domain wall meets the surface will
be associated with a chiral fermion mode.  Like the structure in Fig.
\ref{chiraldevice}(a), this may be interpreted as an edge state on a domain
wall between the ``half quantized" quantum Hall states of the topological
insulator surfaces.  However, an equally valid interpretation is that the
domain wall itself forms a single two dimensional integer quantum Hall state with an
edge state.  Our framework for topologically classifying the line defects
underlies the equivalence between these two points of view.

Mong, Essen and Moore \cite{mong10} have introduced
a {\it different} kind of antferromagnetic topological insulator that relies on
the symmetry of time reversal combined with a lattice translation.  Due to the
necessity of translation symmetry, however, such a phase is not robust to disorder.
They found that chiral Dirac modes occur at certain step edges in such crystals.  These chiral
modes can also be understood in terms of the invariant \eqref{2ndchern}. Note that these chiral modes
survives in the presence of disorder even though the bulk state does not.  Thus, the chiral mode,
protected by the strong invariant \eqref{2ndchern}, is more robust than the
bulk state that gave rise to it.

If one imagines weakening the coupling between the two
antiferromagnetic topological insulators (using our terminology, not that of Mong, et al.\cite{mong10})
and taking them apart, then at some
point the chiral mode has to disappear.  At that point, rather than taking the
``shortest path" between $\pi \pm \epsilon$, $\theta$ takes a path that passes
through $0$.  At the transition between the ``short path" and the ``long path"
regimes, the gap on the domain wall must go to zero, allowing the chiral mode
to escape.  This will have the character of a plateau transition in the 2D
integer quantum Hall effect.

Structures involving magnetic topological insulators would be extremely
interesting to study because with them it is possible to create chiral fermion
states with a single material.  Indeed, one can imagine scenarios where a
magnetic memory, encoded in magnetic domains, could be read by measuring the
electrical transport in the domain wall chiral fermions.

To model the chiral fermions in these structures we begin with the
simple three dimensional model for trivial and topological
insulators considered in Ref. \onlinecite{qihugheszhang08},
\begin{equation}
{\cal H}_0 =  v \mu_x \vec\sigma \cdot {\bf k} + (m+ \epsilon |{\bf k}|^2)
\mu_z.
\label{h0chiral}
\end{equation}
Here $\vec\sigma$ represents spin, and $\mu_z$ describes an orbital degree of freedom.
$m>0$ describes the trivial insulator and $m<0$ describes the
topological insulator.  An interface where $m$ changes sign is
then associated with gapless surface states.

Next consider time reversal symmetry breaking perturbations, which
could arise from exchange fields due to the presence of magnetic
order.  Two possibilities include
\begin{eqnarray}
{\cal H}_{af} = h_{af} \mu_y  \label{haf},\\
{\cal H}_{f} = \vec h_f \cdot \vec\sigma \label{hf}.
\end{eqnarray}
Either  $h_{af}$ or $h_{f,z}$ will introduce a gap in the surface states,
but they have different physical content.
${\cal H}_0$ has an inversion symmetry given by
${\cal H}_0({\bf k}) = P {\cal H}_0(-{\bf k}) P$ with $P = \mu_z$.
Clearly, ${\cal H}_{f}$ respects this inversion symmetry.  ${\cal H}_{af}$ does
not respect $P$, but does respect $P\Theta$.  We therefore associate
${\cal H}_f$ with ferromagnetic order and ${\cal H}_{af}$ with antiferromagnetic order.

Within the adiabatic approximation, the topological invariant
\eqref{2ndchern} can be evaluated in the presence of either
\eqref{haf} or \eqref{hf}.
The antiferromagnetic perturbation \eqref{haf} is most straightforward to analyze because
${\cal H}_0 + H_{af}$ is a combination of 5 anticommuting Dirac matrices.  On
a circle surrounding the junction parameterized by $s$  it can
be written in the general form
\begin{equation}
H({\bf k},s) = {\bf h}({\bf k},s) \cdot \vec\gamma,
\label{hgamma}
\end{equation}
where $\vec\gamma = (\mu_x\sigma_x,\mu_x\sigma_y,\mu_x\sigma_z, \mu_z,\mu_y)$
and  ${\bf h}({\bf k},s) = (v {\bf k},m(s)+ \epsilon|{\bf k}|^2, h_{af}(s))$.
For a model of this form, the second Chern number \eqref{2ndchern} is given
simply by the winding number of the unit vector $\hat{\bf d}({\bf k},s) = {\bf
h}/|{\bf h}|\in S^4$ as a function of ${\bf k}$ and $s$.  This is most
straightforward to evaluate in the limit $\epsilon\rightarrow 0$, where $\hat{\bf
d}$ is confined to the ``equator" $(d_1,d_2,d_3,0,0)$ everywhere except near
${\bf k}\sim 0$ and $|{\bf k}|\gtrsim 1/\epsilon$.  The winding number is
determined by the behavior at ${\bf k} \sim 0$, and may be expressed by
\eqref{n(theta)} with $\theta$ given by
\begin{equation}
e^{i\theta} = \frac{m + ih_{af}}{\sqrt{m^2 + h_{af}^2}}.
\end{equation}

We therefore expect a topological line defect to occur at an intersection
between planes where $m$ and $h_{af}$ change sign.  The chiral fermion mode
associated with this defect can seen explicitly if we solve a
simple linear model, $m = f_z z$, $h_{af} = f_y y$.  This model, which has the
form of a harmonic oscillator, is solved in Appendix \ref{appendix:ZM}, and explicitly gives
the chiral Dirac fermion mode with dispersion
\begin{equation}
E(k_x) = v {\rm sgn}(f_z f_y) k_x.
\label{E(kx)}
\end{equation}

\subsection{Class D: chiral Majorana fermions}
\label{sec:linechiralmajorana}

\subsubsection{Topological invariant}
\label{sec:linechiralmajoranainvariant}

A line defect in a superconductor without time reversal symmetry is
characterized by an integer topological invariant that determines the number of
associated chiral Majorana fermion modes.  Since the BdG Hamiltonian
characterizing a superconductor has the same structure as the Bloch
Hamiltonian, we can analyze the problem by ``forgetting" about the
particle hole symmetry and treating the BdG Hamiltonian as if it was
a Bloch Hamiltonian.  The second Chern number,
given by \eqref{2ndchern} can be defined.  It can be verified that
any value of the Chern number is even under particle-hole symmetry, so that
particle-hole symmetry does not rule out a non zero Chern number.
We may follow the same steps as (\ref{2ndchern}-\ref{theta(s)}) to express the
integer topological invariant as
\begin{equation}
\tilde n = {1\over{8\pi^2}} \int_{T^3\times S^1} {\rm Tr}[\tilde{\cal F} \wedge
\tilde{\cal F}],
\label{tilden}
\end{equation}
where $\tilde{\cal F}$ is the curvature form characterizing the BdG theory.
As in \eqref{n(theta)}, $\tilde n$ may be expressed as a winding number of $\tilde\theta$,
which is expressed as an integral over the Brillouin
zone of the Chern Simons 3 form.  The difference between $n$ and
$\tilde n$ is that $\tilde n$ characterizes a BdG Hamiltonian.  If we
considered the BdG Hamiltonian for a non superconducting insulator, then due to
the doubling in the BdG equation, we would find
\begin{equation}
\tilde n = 2 n.
\end{equation}
In this case, the chiral Dirac fermion that occurs for a $2\pi$ ($n=1$) winding of
$\theta$ corresponds to a $4\pi$ ($n=2$) winding of $\tilde\theta$.
Superconductivity allows for the possibility of a $2\pi$ winding in
$\tilde\theta$: a chiral Dirac fermion can be split into a pair of chiral
Majorana fermions.

\subsubsection{Dislocation in a layered topological superconductor}
\label{sec:linechiralmajoranalayered}

The simplest example to consider is a dislocation in a three dimensional
superconductor.  The discussion closely parallels Section
\ref{sec:qhedislocation}, and we find
\begin{equation}
\tilde n = {1\over{2\pi}} {\bf B} \cdot \tilde {\bf G}_c,
\end{equation}
where ${\bf B}$ is the Burgers vector of the dislocation and $\tilde{\bf G}_c$
characterizes the triad of first Chern numbers characterizing the 3D BdG
Hamiltonian.  A 3D system consisting of layers of a 2D topological
superconductor will be characterized by a non zero $\tilde{\bf G}_c$.
Since, as a 3D superconductor, the layered structure is in the topologically
trivial class, such a state could be referred
to as a {\it weak topological superconductor}.

The simplest model system in this class is a stack of 2D $p_x+ip_y$
superconductors.  A dislocation would then have $\tilde n = 1$ and a single
chiral Majorana fermion branch.
A possible physical realization of the weak topological superconductor
state is Sr$_2$RuO$_4$, which may exhibit triplet $p_x+ip_y$ pairing.
Since the spin up and
spin down electrons make two copies of the spinless state, a dislocation will
be associated with $\tilde n = 2$.  Thus, we predict that there will be two
chiral Majorana modes bound to the dislocation, which is the same as a single
chiral Dirac fermion mode.

\subsubsection{Superconductor Heterostructures}
\label{sec:linechiralmajoranahetero}

\begin{figure}
\epsfxsize=3in
\epsfbox{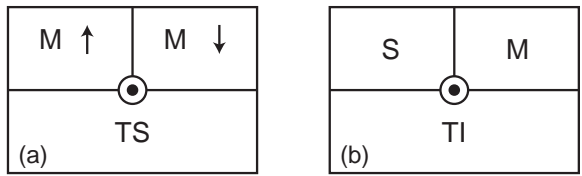}
\caption{Heterostructure geometries for Chiral Majorana fermions.  (a) shows
a magnetic domain wall on the surface of a topological superconductor, while (b)
shows an interface between a superconductor and a magnet on the surface of a
topological insulator.}
\label{mchiraldevice}
\end{figure}

We now consider heterostructures with associated chiral Majorana
modes.  The simplest to consider is a BdG analog of the structures
considered in Fig. \ref{mchiraldevice}.  These would involve, for
example an interface between a 3D time reversal invariant topological superconductor with
a magnetic material with a magnetic domain wall.  The analysis of
such a structure is similar to that in Eq. \ref{h0chiral} if we
replace the Pauli matrices describing the orbital degree of freedom
$\vec \mu$ with Pauli matrices describing Nambu space $\vec\tau$.
Protected chiral Majorana fermion modes of this sort on the surface of
$^3$He-B with a magnetic domain wall have been recently discussed by
Volovik\cite{volovik10}.

In Ref. \onlinecite{fukane08} a different method for engineering chiral Majorana
fermions was introduced by combining an interface between
superconducting and magnetic regions on the surface of a
topological insulator.  To describe this requires the 8 band model
introduced in Ref. \onlinecite{teokane10},
\begin{equation}
H = \tau_z \mu_x \vec\sigma\cdot {\bf k} + (m + \epsilon |{\bf k}|^2) \tau_z\mu_z + \Delta
\tau_x + h \mu_y.
\label{h0d}
\end{equation}
(Here, for simplicity we consider only the antiferromagnetic term).  The
surface of the topological insulator occurs at a domain wall (say, in the x-y plane) where
$m(z)$ changes sign.  The superconducting order parameter $\Delta$
and magnetic perturbation $h$ both lead to an energy gap in the surface states.
This Hamiltonian is straightforward to analyze because $[{\cal
H},\tau_x\mu_y]=0$, which allows the $8\times 8$ problem to be divided into two
$4\times 4$ problems, which have superconducting/magnetic mass terms $\Delta
\pm h$.  Near a defect where $\Delta = h$ the  $\Delta + h$ gap never
closes, while the $\Delta - h$ gap can be critical.
$\Delta>h$ leads to a superconducting state, while $\Delta<h$ leads to a
quantum Hall like state.  There is a transition between the two at $\Delta =
h$.

An explicit model for the line defect can be formulated with $m(z) = f_z z$,
$\Delta - h = f_y y$ and $\Delta + h = M$.  The topological invariant \eqref{tilden} can be
evaluated using a method similar to \eqref{hgamma}, and the chiral Majorana states can
be explicitly solved
along the lines of \eqref{E(kx)}.

\subsection{Class AII: Helical Dirac fermions}
\label{sec:linehelicaldirac}

\subsubsection{Topological Invariant}
\label{sec:linehelicaldiracinvariant}

Line defects for class AII are characterized by a $\mathbb{Z}_2$ topological
invariant.  To develop a formula for this invariant we follow the approach used
in Ref. \onlinecite{fukane06} to describe the invariant characterizing the quantum spin Hall
insulator.

As in the previous section, a line defect in three dimensions is associated with a four
parameter space $({\bf k},{\bf r})\in T^3\times S^1$.  Due to time reversal symmetry, the
second Chern number that characterized the line defects in \eqref{2ndchern} must be zero.  Thus
there is no obstruction to defining Bloch basis functions $|u({\bf k},{\bf r})\rangle$
continuously over the entire base space.  However, the time reversal
relation between $(-{\bf k},{\bf r})$ and $({\bf k},{\bf r})$ allows for an
additional constraint, so that the state is specified by the degrees of freedom
in {\it half} the Brillouin zone.

As in Ref. \onlinecite{fukane06} it is useful to define a matrix
\begin{equation}
w_{mn}({\bf k},{\bf r}) = \langle u_m({\bf k},{\bf r}) | \Theta |u_n(-{\bf
k},{\bf r})\rangle.
\label{wmn(k,r)}
\end{equation}
Because $|u_m({\bf k},{\bf r})\rangle$ and $|u_n(-{\bf k},{\bf r})\rangle$ are
related by time reversal symmetry $w({\bf k},{\bf r})$ is a unitary matrix,
that depends on the gauge choice for the basis functions.
Locally it is possible to choose a basis in which
\begin{equation}
w({\bf k},{\bf r}) = w_0,
\label{trconstraint}
\end{equation}
where $w_0$ is independent of ${\bf k}$ and ${\bf r}$, so that states
at $(\pm {\bf k},{\bf r})$ have a fixed relation.
Since for ${\bf k}=0$ $w = -w^T$, $w_0$ must be antisymmetric.  A natural
choice is thus $w_0 = i \sigma_2 \otimes 1$.

The $\mathbb{Z}_2$
topological invariant is an obstruction to finding such a constrained basis globally.
The constrained basis can be defined on two patches, but the basis functions on
the two patches are necessarily related by a topologically non trivial
transition function.  In this sense, the $\mathbb{Z}_2$ invariant resembles the
second Chern number in \eqref{2ndchern}.

In Appendix \ref{appendix:defectsaiidiii}
we will generalize the argument developed in Ref. \onlinecite{fukane06} to show that
the transition function relating the two patches defines the $\mathbb{Z}_2$
topological invariant, which may be written\cite{fukui09}
\begin{equation}
\nu = {1\over{8\pi^2}}\left(\int_{{1\over 2}T^3 \times S^1} {\rm Tr}[{\cal F}\wedge{\cal
F}]- \int_{\partial{1\over 2}T^3 \times S^1} {\cal Q}_3 \right) \ {\rm mod} \
2,
\label{fukane}
\end{equation}
where ${\cal F}$ and ${\cal Q}_3$ are expressed in terms of the Berry's
connection ${\cal A}$ using \eqref{F(A)} and \eqref{q3}.  The integral is over half of the
base space $(1/2)(T^3\times S^1)$, defined such that $({\bf k},{\bf r})$ and
$(-{\bf k},{\bf r})$ are never both included.  The second term is over the
boundary of $(1/2)(T^3\times S^1)$, which is closed under $({\bf k},{\bf r})
\rightarrow (-{\bf k},{\bf r})$.
Eq. \ref{fukane} must be used with care because
the Chern Simons form in the second term depends on the gauge.   A different
continuous gauge can give a different $\nu$, but due to \eqref{trconstraint},
they must be related by an even integer.  Thus, an odd number is
distinct.

In addition to
satisfying \eqref{trconstraint}, it is essential
to use a gauge in which at least ${\cal Q}_3$ is
continuous on $\partial{1\over 2}T^3 \times S^1$
(though not necessarily on all of ${1\over 2}T^3\times S^1$).  This
continuous gauge can always be found if the base space is a sphere
$S^4$.  However for $T^3\times S^1$, the ``weak" topological
invariants can pose an obstruction to finding a continuous gauge.  We
will show how to work around this difficulty at the end of the
following section.

\subsubsection{Dislocation in a weak topological insulator}
\label{sec:linehelicaldiracdislocation}

Ran, Zhang and Vishwanath recently studied the problem of a line dislocation
in a topological insulator\cite{ran09}.  They found that an insulator with non trivial
weak topological invariants
can exhibit topologically protected helical modes at an appropriate line
dislocation.  In this section we will show that these protected modes are
associated with a non trivial $\mathbb{Z}_2$ invariant in \eqref{fukane}.
In addition to providing an explicit example for this invariant, this
formulation provides additional insight into why protected modes can exist
in a weak topological insulator.  As argued in Ref. \onlinecite{fkm07,fukane07}, the weak topological
invariants lose their meaning in the presence of disorder.  The present
considerations show that the helical modes associated with the dislocation
are protected by the {\it strong} topological invariant associated with the
line defect.  Thus if we start with a perfect crystal and add disorder, then
the helical modes remain, even though the crystal is no longer a weak topological
insulator.  The helical modes remain even if the disorder destroys the
crystaline order, so that dislocations become ill defined, {\it provided}
the mobility gap remains finite in the bulk crystal.  In this case, the
Hamiltonian has a non trivial winding around the line defect, even though
the defect has no obvious structural origin.
Thus, the weak topological insulator provides a {\it route} to
realizing the topologically protected line defect.  But once present, the
line defect is more robust than the weak topological insulator.

To evaluate the $\mathbb{Z}_2$ invariant \eqref{fukane} for a line dislocation
we repeat the analysis in section \ref{sec:qhedislocation}.  Because of the subtlety with
the application of \eqref{fukane} we will first consider the simplest case of a
dislocation in a weak topological insulator.  Afterwards we will discuss the
case of a crystal with both weak and strong invariants.

The Bloch functions on a circle surrounding a dislocation are described by
\eqref{dislocationu},
and the evaluation of ${\rm Tr}[{\cal F}\wedge {\cal F}]$ proceeds exactly as
in (\ref{dislocationa}-\ref{dislocationff}).  To evaluate the second term in
\eqref{fukane} we need the Chern Simons
3 form.  One approach is to use \eqref{q3} and \eqref{dislocationa}.
However, this is not continuously
defined on $\partial (1/2)(T^3\times S^1)$ because ${\cal A}$ has a term
${\bf B}\cdot {\bf k} ds$ that is discontinuous at the Brillouin zone boundary.
An alternative is to write
\begin{equation}
{\cal Q}_3 ={\rm Tr}[ {\bf B}\cdot \left( 2 {\cal A}^0\wedge d{\bf k}
- [{\cal F}^0,{\bf a}^p]\right)\wedge ds].
\label{newq3}
\end{equation}
From \eqref{dislocationff} this
clearly satisfies ${\rm Tr}[{\cal F}\wedge{\cal F}] = d {\cal Q}_3$, and it
is defined continuously on $\partial (1/2)(T^3\times S^1)$ as long as ${\cal
A}^0$ is continuously defined on $\partial (1/2)T^3$.  For a weak topological
insulator this is always possible, provided $(1/2)T^3$ is defined appropriately.
Eq. \ref{newq3} differs from Eq. \ref{q3} by a total derivative.

Combining \eqref{dislocationff}, \eqref{newq3} and
\eqref{fukane}, the terms involving ${\bf a}^p$ cancel because ${\bf
a}^p$ is globally defined.   (Note that ${\bf a}^p$ is unchanged by a ${\bf k}$
dependent -- but ${\bf r}$ independent -- gauge transformation).  This can not be
said of the term involving ${\cal A}^0$, however, because in a weak topological
insulator ${\cal A}^0$ is {\it not} globally defined on $(1/2)T^3$.  Performing
the trivial integral over $s$ we then find
\begin{equation}
\nu = {1\over {2\pi}} {\bf B} \cdot {\bf G}_\nu \ {\rm mod} \ 2,
\label{bdotgnu}
\end{equation}
where
\begin{equation}
{\bf G}_\nu = \int_{{1\over 2}T^3} {\rm Tr}[{\cal F}^0]\wedge d{\bf k}
- \int_{\partial{1\over 2}T^3} {\rm Tr}[{\cal A}^0]\wedge d{\bf k}.
\label{gnu}
\end{equation}

The simplest case to consider is a weak topological insulator consisting of
decoupled layers of 2D quantum spin Hall insulator stacked with a lattice
constant $a$ in the $z$ direction.
In this case ${\cal F}^0 = {\cal F}^0(k_x,k_y)$ is independent of $k_z$, so
the $k_z$ integral can be performed trivially.  This leads to
${\bf G}_\nu = (2\pi/a)\nu \hat {\bf z}$, where
\begin{equation}
\nu =  {i\over{2\pi}}\left[\int_{{1\over 2}T^2} {\rm Tr}[{\cal F}^0] -
\int_{\partial{1\over 2}T^2} {\rm Tr}[{\cal A}^0]\right]
\label{nu2D}
\end{equation}
is the 2D $\mathbb{Z}_2$ topological invariant characterizing the individual
layers.

Eq. \ref{gnu} also applies to a more general 3D weak topological insulator.
A weak topological insulator is characterized by a triad of $\mathbb{Z}_2$
invariants $(\nu_1\nu_2\nu_3)$ that define a mod 2 reciprocal lattice
vector\cite{fkm07,fukane07},
\begin{equation}
{\bf G}_\nu = \nu_1 {\bf b}_1+\nu_2{\bf b}_2 + \nu_3 {\bf b}_3,
\label{gnutopo}\end{equation}
where
${\bf b}_i$ are primitive reciprocal lattice vectors corresponding to
primitive lattice vectors ${\bf a}_i$ (such that ${\bf a}_i\cdot{\bf b}_j =
2\pi \delta_{ij}$).  The indices $\nu_i$ can
be determined by evaluating the 2D invariant \eqref{nu2D} on the time reversal
invariant plane ${\bf k} \cdot {\bf a}_i = \pi$.

To show that ${\bf G}_\nu$ in \eqref{gnu} and \eqref{gnutopo} are
equivalent, consider ${\bf G}_\nu \cdot {\bf a}_1$ in \eqref{gnu}.
If we write ${\bf k} = x_1 {\bf b}_1 + x_2 {\bf b}_2 + x_3{\bf
b}_3$, then the integrals over $x_2$ and $x_3$ have the form of \eqref{nu2D}.
Since this is quantized, it must be independent of $x_1$, and will be given by
its value at $x_1 = 1/2$.  This then gives ${\bf G}_\nu \cdot {\bf a}_1 = 2\pi
\nu_1$.  A similar analysis of the other components establishes the
equivalence.  A non trivial value of Eq. \ref{bdotgnu} is
the same as the criterion for the existence of protected helical modes on a
dislocation Ran, Zhang, Vishwanath\cite{ran09} derived using a different method.

Evaluating \eqref{gnu} in a crystal that is {\it both} a strong topological
insulator and a weak topological insulator (such as Bi$_{1-x}$Sb$_x$) is
problematic because the 2D invariants evaluated on the planes $x_1=0$ and $x_1=1/2$
are necessarily {\it different} in a strong topological insulator.
This arises because a non trivial strong topological invariant $\nu_0$ is an
obstruction to continuously defining ${\cal A}^0$ on $\partial (1/2)T^3$, so
\eqref{nu2D} can not be evaluated continuously between $x_1=0$ and $x_1 = 1/2$.
From the point of view of the topological classification of the {\it defect} on
$T^3 \times S^1$, $\nu_0$ is like a  {\it weak} topological invariant because it a
property of $T^3$ and is independent of the real space parameter $s$ in $S^1$.
Thus this complication is a manifestation of the fact that topological classification of
Hamiltonians on $T^3\times S^1$ has more structure than those on $S^4$.
The problem is not with the existence of the invariant $\nu$ on $T^3 \times
S^1$, but rather with applying the formulas (\ref{fukane},\ref{gnu}).  The problem can be
circumvented with the following trick.

Consider an auxiliary Hamiltonian $\tilde {\cal H}({\bf k},{\bf r}) =
{\cal H}({\bf k},{\bf r}) \oplus {\cal H}_{STI}({\bf k})$, where ${\cal
H}_{STI}$ is a simple model Hamiltonian for a strong topological insulator like
Eq. \ref{h0chiral}, which can be chosen such that it is a constant independent of ${\bf k}$
everywhere except in a small neighborhood close to ${\bf k} = 0$ where a band
inversion occurs.  Adding such a Hamiltonian that is independent of ${\bf r}$
will have no effect on the topologically protected modes associated with a
line defect, so we expect the invariant $\nu$ to be the same for both
${\cal H}({\bf k},{\bf r})$ and $\tilde{\cal H}({\bf k},{\bf r})$.
If ${\cal H}({\bf k},{\bf r})$ has a non trivial strong
topological invariant $\nu_0=1$ then $\tilde{\cal H}({\bf k},{\bf r})$ will
have $\nu_0 = 0$, so that Eq. \ref{gnu} can be applied.   ${\bf G}_\nu$ will
then be given by the 2D invariant \eqref{nu2D} evaluated for $\tilde{\cal H}$,
which will be independent of $x_1$.  Since ${\cal H}_{STI}({\bf k})$ is ${\bf
k}$ independent everywhere except a neighborhood of ${\bf k}=0$, this will
agree with the 2D invariant evaluated for ${\cal H}$ at $x_1 = 1/2$, but
not $x_1=0$.  It then follows that even in a strong topological insulator
the invariant characterizing a line dislocation is
given by \eqref{bdotgnu}, where ${\bf G}_\nu$ is given by \eqref{gnutopo} in
terms of the weak topological invariants.

\subsubsection{Heterostructure geometries}
\label{sec:linehelicaldirachetero}

In principle it may be possible to realize 1D helical fermions in a 3D
system that does not rely on a weak topological insulating state.
It is possible to write down a 3D model, analogous to \eqref{h0chiral} that has bound
helical modes.  However, it is not clear how to physically implement this
model.  This model will appear in a more physical context as a BdG theory in
the following section.

\subsection{Class DIII: helical Majorana fermions}
\label{sec:linehelicalmajorana}

Line defects for class DIII are characterized by a $\mathbb{Z}_2$ topological
invariant that signals the presence or absence of 1D helical Majorana fermion
modes.  As in Section \ref{sec:linechiralmajorana}, the BdG Hamiltonian has the same structure as a
Bloch Hamiltonian, and the $\mathbb{Z}_2$ invariant can be deduced by
``forgetting" the particle hole symmetry, and treating the problem as if it was
a Bloch Hamiltonian in class AII.

There are several ways to realize helical Majorana fermions.
The simplest is to consider the edge of a 2D time reversal invariant superconductor
or superfluid, or equivalently a dislocation in a layered version of that 2D
state.  A second is to consider a
topological line defect in a 3D class DIII topological superconductor or
superfluid.  Such line defects are well known in of $^3$He B\cite{grinevich88,volovik03}
and have recently been revisited in Ref. \onlinecite{qi09,silaev10}.

\begin{figure}
\epsfxsize=1.5in
\epsfbox{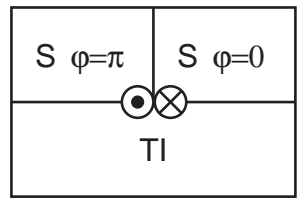}
\caption{Helical Majorana fermions at a linear Josephson junction with phase
difference $\pi$ on the surface of a topological insulator.}
\label{hmdevice}
\end{figure}

Here we will consider a different realization that uses topological
insulators and superconductors.  Consider a linear junction between two
superconductors on the surface of a topological insulator as shown in
Fig. \ref{hmdevice}.  In Ref. \onlinecite{fukane08} it was
shown that when the phase difference between the superconductors is $\pi$ there
are gapless helical Majorana modes that propagate along the junction.  This can
be described by an 8 band minimal model that describes a topological insulator
surface with a superconducting proximity effect,
\begin{equation}
{\cal H} = v \tau_z \mu_x \vec\sigma\cdot {\bf k} + (m + \epsilon|{\bf
k}|^2) \tau_z \mu_z + \Delta_1 \tau_x.
\end{equation}
Here $m$ is the mass describing the band inversion of a topological
insulator, as in \eqref{h0d}, and $\Delta_1$ is the real part of the superconducting gap parameter.
This model has time reversal symmetry with $\Theta = i \sigma_y K$ and
particle-hole symmetry with $\Xi = \sigma_y\tau_y K$.
The imaginary part of the superconducting gap,
$\Delta_2 \tau_y$ violates time reversal symmetry.  A line
junction along the $x$ direction with phase difference $\pi$ at the surface of a topological
insulator corresponds to the intersection of planes where $m(z)$ and $\Delta_1(y)$
change sign.

The $\mathbb{Z}_2$ invariant characterizing such a line defect is
straightforward to evaluate because $[{\cal H},\mu_y\tau_x]=0$.  This
extra symmetry allows a ``spin Chern number" to be defined,
$n_\sigma = (16\pi^2)^{-1}\int {\rm Tr}[\mu_y\tau_x {\cal
F}\wedge{\cal F}]$.  Since the system decouples into two time
reversed versions of \eqref{h0chiral}, $n_\sigma = 1$.  By repeating
the formulation in Appendix \ref{appendix:defectsaiidiii} of the $\mathbb{Z}_2$ invariant $\nu$,
it is straightforward to show that this means $\nu = 1$.

The helical modes can be explicitly seen by solving the linear theory,
$m = f_z z$, $\Delta_1 = f_y y$, which leads to the harmonic
oscillator model studied in Appendix \ref{appendix:ZM}.  In the space of the two
zero modes the Hamiltonian has the form
\begin{equation}
{\cal H} = v k_x \sigma_x,
\end{equation}
and describes 1D helical Majorana fermions.

\subsection{Class C:  Chiral Dirac fermions}
\label{sec:lineclassc}

We finally briefly consider line defects in class C.  Class C can be
realized when time reversal symmetry is broken in a
superconductor without spin orbit interactions that has
even parity singlet pairing.  Line defects are characterized by an
integer topological invariant that determines the number of chiral Majorana
fermion modes associated with the line.  As in class $D$, this may be
evaluated by ``forgetting" the particle-hole symmetry and evaluating
the corresponding Chern number that would characterize class $A$.
The $2\mathbb{Z}$ in Table \ref{linetab} for this case, however, means that the
Chern integer computed in this manner is necessarily {\it even}.
This means that there will necessarily be an even number $2n$ of chiral Majorana
fermion modes, which may equivalently viewed as $n$ chiral Dirac fermion
modes.

An example of such a system would be a 2D $d_{x^2-y^2}+ id_{xy}$ superconductor\cite{laughlin},
which exhibits chiral Dirac fermion edge states, or
equivalently a dislocation in a 3D layered version of that state.

\section{Point defects}
\label{sec:pointdefect}

Point defects can occur at the end of a 1D system ($\delta = 1-0$) or
at topological defects in 2D ($\delta = 2-1$) or 3D ($\delta = 3-2$)
systems.  From the $\delta = 1$ column Table \ref{tab:periodic}, it can be seen that there are five symmetry
classes that can have topologically non trivial point defects.  These
are expected to be associated with protected zero energy bound
states.  Table \ref{pointtab} lists the non trivial classes, along
with the character of the associated zero modes.  In this section we
will discuss each of these cases.

\begin{table}
\centering
\begin{ruledtabular}
\begin{tabular}{ccl}
Symmetry  & Topological classes & $E=0$ Bound States\\
\hline
AIII & $\mathbb{Z}$ &  Chiral Dirac \\
BDI & $\mathbb{Z}$ &  Chiral Majorana \\
D & $\mathbb{Z}_2$ &  Majorana \\
DIII & $\mathbb{Z}_2$ &  Majorana Kramers Doublet  \\
& &  (= Dirac)\\
CII & $2\mathbb{Z}$ &   Chiral Majorana Kramers\\
& &   Doublet (=Chiral Dirac)
\end{tabular}
\end{ruledtabular}
\caption{Symmetry classes supporting non trivial point topological defects
and their associated $E=0$ modes. }
\label{pointtab}
\end{table}

\subsection{Classes AIII, BDI and CII: chiral zero modes}
\label{sec:pointchiral}

\subsubsection{Topological invariant and zero modes}
\label{sec:pointchiralinvariant}

Point defects in classes AIII, BDI and CII
are characterized by integer topological invariants.  The formula for
this integer invariant can be formulated by exploiting the chiral symmetry in
each class.  In a basis where the chiral symmetry operator is $\Pi = \tau_z$,
the Hamiltonian may be written,
\begin{equation}
{\cal H}({\bf k},{\bf r}) = \left(\begin{array}{cc} 0 & q({\bf k},{\bf r}) \\
q({\bf k},{\bf r})^\dagger & 0\end{array}\right).
\label{hchiral}
\end{equation}
When the Hamiltonian has a flattened eigenvalue spectrum ${\cal H}^2 = 1$,
$q({\bf k},{\bf r})$ is a unitary matrix.
For a point defect in $d$
dimensions, the Hamiltonian as a function of $d$ momentum variables and $D=d-1$
position variables is characterized by the winding number associated with the
homotopy $\pi_{2d-1}[U(n\rightarrow\infty)]=\mathbb{Z}$, which is given by
\begin{equation}
n = {(d-1)!\over {(2d-1)! (2\pi i)^d}}\int_{T^d \times S^{d-1}} {\rm Tr}[(q d q^\dagger)^{2d-1}].
\label{nchiral}
\end{equation}
For a Hamiltonian that is built from anticommuting Dirac matrices, ${\cal
H}({\bf k},{\bf r}) = \hat {\bf d}({\bf k},{\bf r}) \cdot \vec \gamma$, this invariant is given
simply by the winding degree of the mapping $\hat {\bf d}({\bf k},{\bf r})$
from $T^{d}\times S^{d-1}$ to $S^{2d-1}$, which is expressed as an integral of
the Jacobian,
\begin{equation}
n = {(d-1)!\over{2\pi^d}}\int_{T^d\times S^{d-1}} d^d{\bf k} d^{d-1}{\bf r} {\partial\hat{\bf
d}({\bf k},{\bf r})\over \partial^d{\bf k}\partial^{d-1}{\bf r}}.
\label{windingdegree}
\end{equation}

In class AIII there are no constraints on $q({\bf k},{\bf r})$ other than
unitarity, so all possible values of $n$ are possible.  There are additional
constraints for the chiral classes with antiunitary symmetries.
As shown in Appendix \ref{appendix:representatives}, this is simplest to see by analyzing the constraints
on the winding degree discussed above.  $n$ must be zero in classes CI and DIII.
There is no constraint on $n$ in class BDI, while $n$ must be even in class
CII.

The topological invariant is related to an index that characterizes the chirality of the zero
modes,
\begin{equation}
n = N_+-N_-,
\label{index}
\end{equation}
where $N_{\pm}$
are the number of zero modes that are eigenstates of $\Pi$ with
eigenvalue $\pm 1$.   To see that these zero modes are indeed protected
consider $N_+=n>0$ and $N_-=0$.  Any term in the Hamiltonian
that could shift any of the $N_+$ degenerate
states would have to have a nonzero matrix element connecting states with the
same chirality.  Such terms are forbidden, though, by the chiral symmetry
$\{{\cal H},\Pi\}=0$.
In the superconducting classes BDI and CII the zero energy states are Majorana
bound states.  In class CII, however, since time reversal symmetry requires
that $n$ must be even, the paired
Majorana states can be regarded as zero energy Dirac fermion states.

In the special case where ${\cal H}({\bf k},{\bf r})$ has the form of
a massive Dirac Hamiltonian, by introducing a suitable regularization for
$|{\bf k}|\rightarrow\infty$ the topological invariant
(\ref{nchiral},\ref{windingdegree}) can be
expressed in a simpler manner as a topological invariant characterizing
the mass term.  In the following subsections we consider this in the
three specific cases $d = 1,2,3$.

\subsubsection{Solitons in d=1}
\label{sec:pointchirald=1}

The simplest topological zero mode occurs in the Jackiw Rebbi model\cite{jackiwrebbi}, which is
closely related to the Su Schrieffer Heeger model\cite{ssh}.  Consider
\begin{equation}
{\cal H}(k,x) = v k \sigma_x  + m \sigma_y.
\end{equation}
Domain walls where $m(x)$ changes sign as a function of $x$ are
associated with the well known zero energy soliton states.

To analyze the topological class requires a regularization for $|k|\rightarrow
\infty$.  This can either be done with a lattice, as in the Su, Schrieffer,
Heeger model or by adding a term $\epsilon k^2 \sigma_y$, as in
\eqref{h0chiral} so that $|k|\rightarrow\infty$ can be replaced by a single point.
In either case, the invariant \eqref{nchiral} changes by $1$
when $m$ changes sign.

\subsubsection{Jackiw Rossi Model in d=2}
\label{sec:pointchirald=2}

Jackiw and Rossi introduced a two dimensional model that has protected zero
modes\cite{jackiwrossi}.  The Hamiltonian can be written
\begin{equation}
{\cal H}({\bf k},{\bf r}) = v \vec \gamma \cdot {\bf k}
+ \vec\Gamma\cdot\vec\phi({\bf r}),
\label{jackiwrossi}
\end{equation}
where ${\bf k} = (k_x,k_y)$ and
$(\gamma_1,\gamma_2)$ and $(\Gamma_1,\Gamma_2)$
are anticommuting Dirac matrices.
They showed that the core of a vortex where $\phi = \phi_1 + i \phi_2$ winds by
$2\pi n$ is associated with $n$ zero modes that are protected by the chiral
symmetry.  Viewed as a BdG Hamiltonian, these zero modes are Majorana bound states.

This can be interpreted as a Hamiltonian describing superconductivity
in Dirac fermions.  In this interpretation the Dirac matrices are
expressed as $(\gamma_1,\gamma_2) = \tau_z(\sigma_x,\sigma_y)$ and
$(\Gamma_1,\Gamma_2) = (\tau_x,\tau_y)$, where $\vec\sigma$ is a Pauli
matrix describing spin and $\vec\tau$ describes particle-hole space.
The superconducting pairing term is $\Delta = \phi_1 + i\phi_2$.
In this interpretation a
vortex violates the physical time reversal symmetry $\Theta = i\sigma_y K$.  However,
even in the presence of a vortex this model has a fictitious ``time reversal symmetry"
$\tilde\Theta = \sigma_x\tau_x K$ which satisfies $\tilde\Theta^2 = +1$.  This
symmetry would be violated by a finite chemical potential term $\mu \tau_z$.
Combined with particle-hole symmetry
$\Xi = \sigma_y\tau_y K$ ($\Xi^2 = +1$), $\tilde\Theta$ defines the
BDI class with chiral symmetry $\Pi = \sigma_z\tau_z$.

Evaluating the topological invariant \eqref{nchiral} again requires a
$|{\bf k}|\rightarrow\infty$ regularization.
One possibility
is to add $\epsilon |{\bf k}|^2 \tau_x$, so that $|{\bf
k}|\rightarrow \infty$ can be replaced by a single point.  In this
case the invariant can be determined by computing the winding degree
of $\hat {\bf d}({\bf k},{\bf r})$ on $S^3$.
In the limit $\epsilon\rightarrow 0$ the ${\bf k}$ integral can
be performed, so that \eqref{nchiral} can be expressed as the winding number of the phase
of $\phi_1+i\phi_2 = |\Delta|e^{i\varphi}$,
\begin{equation}
n = {1\over{2\pi }} \int_{S^1} d\varphi.
\label{phiwinding}
\end{equation}

\subsubsection{Hedgehogs in $d=3$}
\label{sec:pointchirald=3}

In Ref. \onlinecite{teokane10} we introduced a three dimensional model for Majorana bound
states that can be interpreted as a theory of a vortex at the
interface between a superconductor and a topological insulator.  In
the special case that the chemical potential is equal to zero, model
has the same form as \eqref{jackiwrossi}, except that now all of the
vectors are three dimensional.  In the topological insulator model
we have  $\vec\gamma = (\gamma_1,\gamma_2,\gamma_3) = \mu_x\tau_z \vec\sigma$
and $\vec\Gamma = (\Gamma_1,\Gamma_2,\Gamma_3) = (\mu_z\tau_z,\tau_x,\tau_y)$.
$\vec\tau$ and $\vec\sigma$ are defined as before, while $\vec\mu$
describes a orbital degree of freedom.
The chiral symmetry, $\Pi = \mu_y\tau_z$ is violated if a chemical potential term
$\mu \tau_z$ is included.

Following the same steps that led to \eqref{phiwinding} the invariant
\eqref{nchiral} is given by the winding number of
$\hat\phi = \vec\phi/|\vec\phi|$ on $S^2$,
\begin{equation}
n = {1\over{4\pi}}\int_{S^2} \hat \phi \cdot (d\hat\phi \times
d\hat\phi).
\end{equation}

\subsection{Class D: Majorana bound states}
\label{sec:pointmajorana}

\subsubsection{Topological invariant}
\label{sec:pointmajoranainvariant}

Point defects in class D are characterized by a $\mathbb{Z}_2$ topological
invariant that determines the presence or absence of a Majorana bound state
associated with the defect.   These include the well known end states in a 1D p
wave superconductor and vortex states in a 2d $p_x+ip_y$ superconductor.  In
Ref. \onlinecite{teokane10} we considered such zero modes in a three dimensional BdG theory
describing Majorana zero modes in topological insulator structures.  Here we
develop a unified description of all of these cases.

For a point defect in $d$ dimensions, the Hamiltonian depends on $d$ momentum
variables and $D=d-1$ position variables.  In appendix \ref{appendix:pointdefectclassd}
we show that the $\mathbb{Z}_2$ invariant is given by,
\begin{equation}
\nu = {2\over{d!}}\left({i\over {2\pi}}\right)^d \int_{T^d\times S^{d-1}} {\cal Q}_{2d-1} \ {\rm mod} \ 2,
\label{muchernsimons}
\end{equation}
where ${\cal Q}_{2d-1}$ is the Chern Simons form.
The specific cases of interest are,
\begin{eqnarray}
{\cal Q}_1 &=& {\rm Tr}[{\cal A}],\\
{\cal Q}_3 &=& {\rm Tr}[{\cal A} d {\cal A} + {2\over 3}{\cal A}^3], \label{cs3d}\\
{\cal Q}_5 &=& {\rm Tr}[{\cal A} (d{\cal A})^2 + {3\over 2}{\cal A}^3 d{\cal A} +
{3\over 5}{\cal A}^5].
\end{eqnarray}

It is instructive to see that \eqref{muchernsimons} reduces to \eqref{nchiral}
 in the case in which a system
also has particle-hole symmetry.  In this case, as detailed in Appendix
\ref{appendix:pointdefectclassd}
it is possible to choose a
gauge in which ${\cal A} = q^\dagger dq/2$, so that ${\cal Q}_{2d-1} \propto
(qdq^\dagger)^{2d-1}$.

\subsubsection{End States in a 1D superconductor}
\label{sec:pointmajorana1d}

The simplest example of a point defect in a superconductor occurs in Kitaev's
model\cite{kitaev00} of a one dimensional p wave superconductor.  This is described by a
simple 1D tight binding model for spinless electrons, which includes a nearest
neighbor hopping term $t c_i^\dagger c_{i+1} + {\rm h.c.}$ and a nearest
neighbor p wave pairing term $\Delta c_i c_{i+1} + {\rm h.c.}$.  The
Bogoliubov de Gennes Hamiltonian can then be written as
\begin{equation}
{\cal H}(k) = ( t \cos k - \mu )\tau_z  + \Delta \sin k \tau_x.
\label{1DKitaev}
\end{equation}
This model exhibits a weak pairing phase for $|\mu| < t$ and a strong pairing
phase for $|\mu|>t$.   The weak pairing phase will have zero energy Majorana
states at its ends.

The topological invariant \eqref{muchernsimons} can be easily
evaluated.  We find ${\cal A} = d\theta/2$, where $\theta$ is the polar angle
of ${\bf d}(k) = ( t\cos k - \mu,\Delta \sin k)$.  It follows that for
$|\mu|<t$, the topological invariant is $\nu = 1$ mod 2.

\subsubsection{Vortex in a 2D topological superconductor}
\label{sec:pointmajorana2d}

In two dimensions, a Majorana bound state occurs at a vortex in a
topological superconductor.  This can be easily seen by considering
the edge states of the topological superconductor in the presence of
a hole\cite{readgreen}.  Particle-hole symmetry requires that the quantized
edge states come in pairs.  When the flux is an odd multiple of $h/2e$,
the edge states are quantized such that a zero mode is present.  In
this section we will evaluate the topological invariant
\eqref{muchernsimons} associated with a loop surrounding the vortex\cite{thankroman}.

We begin with the class D BdG Hamiltonian ${\cal H}^0_p(k_x,k_y)$
characterizing the topological superconductor when the superconducting
phase is zero.  We include the
subscript $p$ to denote the first Chern number that classifies the
topological superconductor.  We can then introduce a nonzero
superconducting phase by a gauge transformation,
\begin{equation}
{\cal H}_p({\bf k},\varphi) = e^{-i\varphi\tau_z/2} {\cal H}_p^0({\bf k})
e^{i\varphi\tau_z/2},
\label{phaserotation}
\end{equation}
where $\tau_z$ operates in the Nambu particle-hole space.
We now wish to evaluate \eqref{muchernsimons} for this Hamiltonian
when phase $\varphi(s)$ winds around a vortex.
There is, however, a difficulty because the Chern Simons formula
requires a gauge that is continuous throughout the entire base space
$T^2 \times S^1$.  The nonzero Chern number $p$ characterizing ${\cal
H}_p^0({\bf k})$ is an obstruction to constructing such a gauge.  A
similar problem arose in Section \ref{sec:linehelicaldiracdislocation}, when we
discussed a line dislocation in a weak topological superconductor.
We can adapt the trick we used there to get around the present
problem.  We thus double the Hilbert space to include two copies of
our Hamiltonian--one with Chern number $p$ and one with Chern number $-p$,
\begin{equation}
\tilde{\cal H}^0({\bf k}) = \left(\begin{array}{cc} {\cal H}^0_p({\bf k}) & 0 \\ 0 & {\cal H}^0_{-p}({\bf k})
\end{array}\right).
\end{equation}
We then put the vortex in only the $+p$ component,
\begin{equation}
\tilde{\cal H}({\bf k},\varphi) = e^{-i \varphi q} \tilde{\cal
H}^0({\bf k}) e^{i \varphi q},
\end{equation}
where
\begin{equation}
q = {{1+\tau_z}\over 2} \left(\begin{array}{cc} 1 & 0 \\ 0 & 0
\end{array}\right).
\label{qequation}
\end{equation}
We added an extra phase factor by replacing $\tau_z$ by $1+\tau_z$ in order to
make $e^{i \varphi q}$ periodic under $\varphi\rightarrow\varphi+2\pi$.

Since the Chern number characterizing $\tilde {\cal H}^0({\bf k})$ is
zero, there exists a continuous gauge,
\begin{equation}
|\tilde u_i({\bf k},\varphi)\rangle = e^{i\varphi q} |\tilde
u^0_i({\bf k})\rangle,
\end{equation}
which allows us to evaluate
the Chern Simons integral.  The Berry's connection
$\tilde{\cal A}_{ij} = \langle \tilde u_i |d\tilde u_j\rangle$ is given by
\begin{equation}
\tilde{\cal A} = \tilde{\cal A}^0 + i Q d\varphi,
\end{equation}
where $\tilde{\cal A}^0({\bf k})$ is the connection describing $\tilde {\cal
H}^0({\bf k})$ and $Q_{ij}({\bf k}) = \langle \tilde u^0_i({\bf k})|q|\tilde u^0_j({\bf k})\rangle$.
Inserting this into \eqref{cs3d} and rearranging
terms we find
\begin{equation}
{\cal Q}_3 = {\rm Tr}[ 2 Q \tilde{\cal F}^0 - d(Q \tilde{\cal A}^0) ] \wedge
d\varphi,
\end{equation}
where $\tilde {\cal F}^0 = d\tilde{\cal A}^0 + \tilde{\cal A}^0\wedge\tilde{\cal A}^0$.
Since the second term is a total derivative it can be discarded.  For
the first term there are two contributions from the $1$ and the
$\tau_z$ in \eqref{qequation}.  Upon integrating over ${\bf k}$, the $\tau_z$
term can be shown to vanish as a consequence of particle-hole
symmetry.  The $1$ term simply projects out the Berry curvature of
the original Hamiltonian ${\cal H}^0_p({\bf k})$, so that
\begin{equation}
{\cal Q}_3 = {\rm Tr}[{\cal F}^0]\wedge d\varphi.
\end{equation}
It follows from \eqref{muchernsimons} that the $\mathbb{Z}_2$
invariant characterizing the vortex is
\begin{equation}
\nu = p m \ {\rm mod} \ 2,
\end{equation}
where $p$ is the Chern number characterizing the topological
superconductor and $m$ is the phase winding number associated with
the vortex.

It is also instructive to consider this invariant in the context of
the simple two band model introduced by Read and
Green\cite{readgreen}.  This can be written as a simple tight binding model
\begin{equation}
{\cal H}^0(k_x,k_y) = (t (\cos k_x + \cos k_y) - \mu) \tau_z + \Delta (\sin k_x \tau_x +
\sin k_y \tau_y).
\label{readgreenh}
\end{equation}
where the superconducting order parameter $\Delta$ is real.  As in
\eqref{1DKitaev}, this model exhibits weak and strong pairing phases for
$|\mu|<t$ and $|\mu|>t$.  These are distinguished by the Chern
invariant, which in turn is related to the winding number on $S^2$ of the
unit vector $\hat {\bf d}({\bf k})$, where $\vec d({\bf k})$ are the
coefficients of $\vec\tau$ in \eqref{readgreenh}.
A nonzero superconducting phase is again introduced by rotating about
$\tau_z$ as in \eqref{phaserotation}.  Here we wish to show that in this
two band model the
$\mathbb{Z}_2$ invariant $\nu$ can be understood from a geometrical
point of view.

\begin{figure}
\epsfxsize=3.3in
\epsfbox{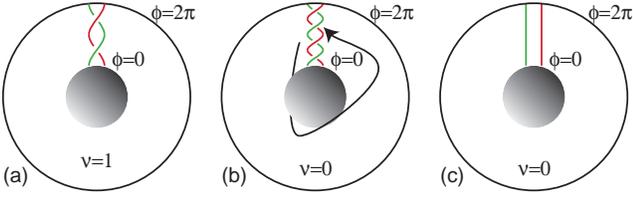}
\caption{Visualization of the $\mathbb{Z}_2$ Pontrjagin invariant characterizing maps from
$({\bf k},\phi) \in S^2 \times S^1$ to $S^2$ when the winding degree for each $\phi$ is 1.
The inner sphere corresponds to ${\bf k}$ at $\phi=0$, while the outer sphere is $\phi = 2\pi$.
The lines depict inverse images of two specific points on $S^2$, which are lines connecting
the inner and outer spheres.  In (a) they have one twist, which can not be eliminated.  The double
twist in (b) can be unwound by smoothly dragging the paths around the sphere to arrive at (c),
which has no twist. }
\label{pontrjagin}
\end{figure}

The $Z_2$ invariant characterizing a vortex
can be understood in terms of the topology of the maps $\hat {\bf
d}(k_x,k_y,\phi)$.
from $T^2 \times S^1$ to $S^2$.  These maps were first classified by
Pontrjagin\cite{pontrjagin}, and have also appeared in other physical
contexts\cite{jaykka,kapitanski,teokane10}.
Without losing generality, we can reduce the torus $T^2$ to a sphere
$S^2$, so the mappings are $S^2 \times S^1 \rightarrow S^2$.
When for  fixed $\phi$ $\hat {\bf d}(k_x,k_y,\phi)$ has an $S^2$ winding
number of $\pm p$, the topological classification is
$\mathbb{Z}_{2p}$.  In the case of interest, $p=1$, so there are two
classes.

This $\mathbb{Z}_2$ Pontrjagin invariant can be understood
pictorially by considering inverse image paths in $({\bf k},\phi)$
space, which map to two specific points on $S^2$.  These correspond
to 1D curves in $S^2 \times S^1$.    Fig. \ref{pontrjagin} shows
three examples of such curves.  The inner sphere corresponds to
$\phi=0$, while the outer sphere corresponds to $\phi=2\pi$.  Since
$p=1$, for every point on $S^2$ the inverse image path is a single
curve connecting the inner and outer spheres.  The key point is to examine the
linking properties of these curves.  The $\mathbb{Z}_2$ invariant
describes the number of twists in a pair of inverse image paths,
which is $1$ in (a), $2$ in (b) and $0$ in (c).  The configuration in (b)
can be continuously deformed into that in (c) by dragging the paths
around the inner sphere.  This can be verified by a simple
demonstration using your belt.  The twist in (a), however, can not be undone.
The number of twists thus defines the $\mathbb{Z}_2$ Pontrjagin
invariant.

\subsubsection{Superconductor Heterostructures}
\label{sec:pointmajoranahetero}

Finally, in three dimensions, a non trivial point defect can occur at
a superconductor heterostructure.  An example is a vortex in the
superconducting state at the interface between a superconductor and a
topological insulator.  As shown in Ref. \onlinecite{teokane10}, this can be described by
the simple Hamiltonian,
\begin{equation}
{\cal H} = v \tau_z \mu_x \vec\sigma \cdot {\bf k} - \mu \tau_z + (m +
\epsilon|{\bf k}|^2 ) \tau_z \mu_z + \Delta_1 \tau_x + \Delta_2
\tau_y.
\end{equation}
Here $m$ is a mass which distinguishes a topological insulator from a
trivial insulator, and $\Delta = \Delta_1 + i\Delta_2$ is a
superconducting order parameter.  For $\mu = 0$, this Hamiltonian has
the form of the three dimensional version of \eqref{jackiwrossi}
discussed in IV.A.4, where the mass term is characterized by the
vector $\vec\phi = (m,\Delta_1,\Delta_2)$.  A vortex in $\Delta$ at
the interface where $m$ changes sign then corresponds to a hedgehog
singularity in $\vec\phi$.  From \eqref{windingdegree}, it can be seen that the class BDI $\mathbb{Z}$
invariant
is $n=1$.  This then establishes that the class D $\mathbb{Z}$ invariant is $\nu =
1$.  The $\mathbb{Z}_2$ survives when a nonzero chemical potential
reduces the symmetry from class BDI to class D.

\subsection{Class DIII: Majorana doublets}
\label{sec:pointdiii}

Point defects in class DIII are characterized by a $\mathbb{Z}_2$
topological invariant.  These are associated with zero modes, but
unlike class D, the zero modes are required by Kramers theorem to be
doubly degenerate.  The zero modes thus form a Majorana doublet,
which is equivalent to a single Dirac fermion.

In Table \ref{tab:periodic}, Class DIII, $\delta = 1$ is an entry that is similar to Class
AII, $\delta = 2$.  The $\mathbb{Z}_2$ for DIII invariant bears a resemblance to the
invariant for AII, which is a generalization of the $\mathbb{Z}_2$ invariant
characterizing the 2D quantum spin Hall insulator.  In Appendix \ref{appendix:pdDIII} we will
establish a formula that employs the same gauge constraint,
\begin{equation}
w({\bf k},{\bf r}) = w_0,
\label{DIIIgc}
\end{equation}
where $w_0$ is a constant independent of ${\bf k}$ and ${\bf r}$.
$w({\bf k},{\bf r})$ relates the time reversed states at ${\bf k}$ and
$-{\bf k}$ and is given by \eqref{wmn(k,r)}.  {\it Provided} we choose a gauge
that satisfies this constraint, the $\mathbb{Z}_2$ invariant is given by,
\begin{equation}
\tilde\nu=\frac{1}{d!}\left(\frac{i}{2\pi}\right)^d
\int_{T^d\times S^{d-1}}\mathcal{Q}_{2d-1}\quad\mbox{mod 2}.
\label{diiiformula1}
\end{equation}
This formula is almost identical to the formula for a point defect in class D,
but they differ by an important factor of two.  Due to the combination of time reversal
and particle-hole symmetry the Chern Simons integral \eqref{diiiformula1} is
guaranteed to be an integer, but the integer is not gauge invariant.  When the time reversal
constraint is satisfied, the parity $\tilde\nu$ is gauge invariant.  It
then follows that the class D invariant in \eqref{muchernsimons}, $\nu = 0$ mod 2.

In the special case $d=1$ there is a formula that does not rely on the gauge
constraint, though it still requires a globally defined gauge.
It is related to the similar ``fixed point" formula for the
invariant for the 2D quantum spin Hall insulator\cite{fukane06}, and has recently been
employed by Qi, Hughes and Zhang\cite{qhz10} to classify one dimensional time reversal
invariant superconductors.   In class DIII, it is possible to choose a basis
in which the time reversal and particle hole operators are given by
$\Theta = \tau_y K$ and $\Xi = \tau_x K$, so that the chiral operator is
$\Pi = \tau_z$.  In this basis, the Hamiltonian has the form
\eqref{hchiral}, where $q({\bf k},{\bf r}) \rightarrow q(k)$
satisfies $q(-k) = - q(k)^T$.
Thus, ${\rm Pf}[q(k)]$ is defined for the time reversal
invariant points $k = 0$ and $k=\pi$.  $q(k)$ is related to $w(k)$ because in a
particular gauge it is possible to choose $w(k) = q(k)/\sqrt{|{\rm Det}[q(k)]|}$.
The $\mathbb{Z}_2$ invariant is then given by,
\begin{equation}
(-1)^{\tilde\nu} = {{\rm Pf}[q(\pi)]\over{{\rm Pf}[q(0)]}} {\sqrt{{\rm
Det}[q(0)]}\over\sqrt{{\rm Det}[q(\pi)]}},
\label{diiiformula2}
\end{equation}
where the branch $\sqrt{{\rm Det}[q(k)]}$ is chosen continuously
between $k=0$ and $k=\pi$.
The equivalence of \eqref{diiiformula1} and \eqref{diiiformula2} for $d=1$ is
demonstrated in Appendix \ref{appendix:fixedpointformulas}.  Unlike \eqref{diiiformula1}, however,
the fixed point formula \eqref{diiiformula2} does not have a natural generalization for
$d>1$.

\begin{figure}
\epsfxsize=2.5in
\epsfbox{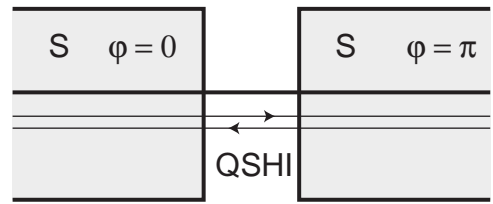}
\caption{A Josephson junction in proximity with the helical edge states of a quantum
spin Hall insulator.  When the phase difference is $\pi$, there is a zero energy Majorana
doublet at the junction.}
\label{pijunction}
\end{figure}

Majorana doublets can occur at topological defects in time reversal
invariant topological superconductors, or in Helium 3B.  Here we consider a different
configuration at a Josephson junction at the edge of a quantum spin
Hall insulator (Fig. \ref{pijunction}).  When the phase difference across the Josephson
junction is $\pi$, it was shown in Ref. \onlinecite{kitaev00,fukane09a} that there is a level
crossing in the Andreev bound states at the junction.  This
corresponds precisely to a Majorana doublet.

This can be described by a the simple continuum 1D theory introduced in Ref. \onlinecite{fukane09a}.
\begin{equation}\label{1Djj}
{\cal H} = v k \tau_z \sigma_z + \Delta_1 \tau_x
\end{equation}
Here $\sigma_z$ describes the spin of the quantum spin Hall edge
state, and $\Delta_1$ is the real superconducting order parameter.
This model has particle-hole symmetry $\Xi = \sigma_y\tau_y K$ and
time reversal symmetry $\Theta = i \sigma_y K$ and is in class DIII.
A $\pi$ junction corresponds to a domain wall where $\Delta_1$
changes sign.  Following appendix \ref{appendix:ZM}, it is straightforward
to see that this will involve a degenerate pair of zero modes
indexed by the spin $\sigma_z$ and chirality $\tau_y$ constrained by
$\tau_y\sigma_z=-1$.

The Hamiltonian \eqref{1Djj} should be viewed as an a low energy
theory describing the edge of a 2D quantum spin Hall insulator.
Nonetheless, we may describe a domain wall where $\Delta_1$ changes
sign using an effective one dimensional theory by introducing a
regularization replacing $\Delta_1$ by $\Delta_1 + \epsilon k^2$.
This regularization will not effect the topological structure of a
domain wall where $\Delta_1$ changes sign. A
topologically equivalent lattice version of the theory then
has the form,
\begin{equation}
{\cal H} = t \sin k \tau_z \sigma_z + (\Delta_1 + u(1-\cos k))\tau_x.
\label{hdiiilattice}
\end{equation}
where we assume $|\Delta_1|< 2u$.

The topological invariant can be evaluated using either
\eqref{diiiformula1} or \eqref{diiiformula2}.  To use
\eqref{diiiformula1}, note that \eqref{hdiiilattice} has exactly the
same form as two copies (distinguished by $\sigma_z=\pm 1$) of
\eqref{1DKitaev}.  The evaluation of \eqref{diiiformula1} then
proceeds along the same lines.  It is straightforward to check that
in a basis where the time reversal constraint \eqref{DIIIgc} is
satisfied (this fixes the relative phases of the $\sigma_z=\pm 1$
states), ${\cal A} = d\theta$, where $\theta$ is the polar angle of
${\bf d}(k) = (t\sin k, \Delta_1 + u(1-\cos k))$.  It follows that a defect
where $\Delta_1$ changes sign has $\tilde\nu = 1$.

To use \eqref{diiiformula2}, we transform to a basis in which
$\Theta = \tau_y K$, $\Xi = \tau_x K$ and $\Pi = \tau_z$.  This is
accomplished by the unitary
transformation $U = \exp[i (\pi/4) \sigma_y\tau_z] \exp[i (\pi/4) \tau_x]$.  Then,
${\cal H}$ has the form of Eq. \ref{h0chiral} with $q(k) = -i( t\sin k \sigma_z
+ (\Delta_1 + u (1-\cos k)) \sigma_y$.
It follows that ${\rm det}[q(k)]$ is real and positive for all $k$.
Moreover,  ${\rm Pf}[q(0)]/\sqrt{{\rm det}[q(0)]} = {\rm sgn}[\Delta_1]$ while
${\rm Pf}[q(\pi)]/\sqrt{{\rm det}[q(\pi)]} =1$.  Again, a defect
where $\Delta_1$ changes sign has $\tilde\nu = 1$.

\section{Adiabatic pumps}
\label{sec:pump}

In this section we will consider time dependent
Hamiltonians ${\cal H}({\bf k},{\bf r},t)$, where in additional to
having adiabatic spatial variation ${\bf r}$ there is a cyclic adiabatic
temporal variation parameterized by $t$.  We will focus on point
like spatial defects, in which the dimensions of ${\bf k}$ and ${\bf
r}$ are related by $d-D = 1$.

Adiabatic cycles in which ${\cal
H}({\bf k},{\bf r},t=T) = {\cal H}({\bf k},{\bf r},t=0)$ can be
classified topologically by considering $t$ to be an additional
``spacelike" variable, defining $\tilde D = D + 1$.  Such cycles will
be classified by the $\delta = 0$ column of Table \ref{tab:periodic}.
Topologically non trivial cycles correspond to adiabatic
pumps.   Table \ref{pumptab} shows the
symmetry classes which host non trivial pumping cycles, along with
the character of the adiabatic pump.  There are two general cases.
Classes A, AI and AII define a charge pump, where after one cycle an
integer number of charges is transported towards or away from the point defect.
Classes BDI and D define a fermion parity pump.  We will discuss
these two cases separately.

We note in passing that the $\delta = 0$ column of Table \ref{tab:periodic} also applies
to topological {\it textures}, for which $d=D$.  For example a spatially dependent
three dimensional band structure ${\cal H}({\bf k},{\bf r})$ can have
topological textures analogous to Skyrmions in a 2D magnet.  Such
textures have recently been analyzed by Ran, Hosur and Vishwanath\cite{ran10b}
for the case of class D,
where they showed that the $\mathbb{Z}_2$ invariant characterizing
the texture corresponds to the fermion parity associated with the
texture.  Thus, non trivial textures are fermions.

\begin{table}
\centering
\begin{ruledtabular}
\begin{tabular}{ccl}
Symmetry  & Topological classes & Adiabatic Pump\\
\hline
A & $\mathbb{Z}$ &  Charge \\
AI & $\mathbb{Z}$ &  Charge \\
BDI & $\mathbb{Z}_2$ &  Fermion Parity \\
D & $\mathbb{Z}_2$ &  Fermion Parity  \\
AII & $2\mathbb{Z}$ &  Charge Kramers Doublet\\
\end{tabular}
\end{ruledtabular}
\caption{Symmetry classes that support non trivial charge or
fermion parity pumping cycles.   }
\label{pumptab}
\end{table}

\subsection{Class A, AI, AII:  Thouless Charge Pumps}
\label{sec:pumpcharge}

The integer topological invariant
characterizing a pumping cycle in class A is simply the Chern
number characterizing the Hamiltonian ${\cal H}({\bf
k},{\bf r},t)$\cite{thouless,thoulessniu}.  Imposing time reversal symmetry has only a minor
effect on this.  For $\Theta^2 = -1$ (Class AII), an odd Chern number violates
time reversal symmetry, so that only even Chern numbers are allowed.
This means that the pumping cycle can only pump Kramers pairs of
electrons.  For $\Theta^2 = +1$ (Class AI) all Chern numbers are consistent with
time reversal symmetry.

The simplest charge pump is the 1D model introduced by Thouless\cite{thouless}.  A
continuum version of this model can be written in the form,
\begin{equation}
{\cal H}(k,t) = v k \sigma_z + (m_1(t)+ \epsilon k^2) \sigma_x + m_2(t)
\sigma_y.
\end{equation}
When the masses undergo a cycle such that the phase of $m_1+im_2$
a single electron is transmitted down the wire.  In this case, ${\cal
H}(k,t)$ has a non zero first Chern number.
The change in the charge associated with a point in a 1D system is
given by the difference in the Chern numbers associated with either
side of the point.  Thus, after a cycle a charge $e$ accumulates at
the end of a Thouless pump.

A two dimensional version of the charge pump can be developed based on Laughlin's
argument\cite{laughlin81} for the integer quantum Hall effect.  Consider a 2D $\nu=1$ integer
quantum Hall state and change the magnetic flux threading a hole from
0 to $h/e$.  In the process, a charge $e$ is pumped to the edge
states surrounding the hole.  This pumping process can be
characterized by the second Chern number characterizing the 2D Hamiltonian ${\cal
H}(k_x,k_y,\theta,t)$, where $\theta$ parameterizes a circle
surrounding the hole.   A similar pump in 3D can be considered, and
is characterized by the third Chern number.

\subsection{Class D, BDI: Fermion Parity Pump}
\label{sec:pumpparity}

Adiabatic cycles of point defects in class D and BDI are
characterized by a $\mathbb{Z}_2$ topological invariant.
In this section we will argue that a non trivial pumping cycle transfers a unit
of fermion parity to the point defect.  This is intimately related to
the Ising non-Abelian statistics associated with defects supporting
Majorana bound states.

Like the point defect in class DIII ($\delta = 1$), the temporal pump ($\delta = 0$)
in class D occupies an entry in Table \ref{tab:periodic} similar to the line
defect ($\delta = 2$) in class AII, so we expect a formula that is similar to the
formula for the 2D quantum spin Hall insulator.  This is indeed the
case, though the situation is slightly more complicated.
The Hamiltonian ${\cal H}({\bf k},{\bf r},t)$ is defined on a base
space $T^d \times S^{d-1}\times S^1$.  In appendix \ref{appendix:pumpinvariant} we will show
that the invariant can be written in a form that resembles
\eqref{fukane},
\begin{equation}
\label{FPPFK}
\nu=\frac{i^d}{d!(2\pi)^d}
 \left[\int_{\mathcal{T}_{1/2}}\mbox{Tr}(\mathcal{F}^d) -
 \oint_{\partial\mathcal{T}_{1/2}}\mathcal{Q}_{2d-1}\right]\quad\mbox{mod
 2},
\end{equation}
where $\mathcal{T}_{1/2}$ is half of the base manifold, say $k_1\in[0,\pi]$,
and the Chern-Simons form $\mathcal{Q}_{2d-1}$ is generated by a continuous
valence frame $u_v({\bf k},{\bf r},t)|_{k_1=0,\pi}$ that obeys certain
particle-hole gauge constraint.  This is more subtle than the time reversal
gauge condition \eqref{trconstraint} for line defects in AII and point defects in DIII.
Unlike \eqref{trconstraint}, we do not have a computational way of checking whether or not
a given frame satisfies the constraint.  Nevertheless, it can be
defined, And in certain simple examples,
the particle-hole constraint is automatically satisfied.

The origin of the difficulty is that unlike time reversal symmetry,
particle hole symmetry connects the conduction and valence bands.  The
gauge constraint therefore involves both.
Valence and conduction frames can be combined to form a unitary
matrix,
\begin{equation}
G_{{\bf k},{\bf r},t}=\left(\begin{array}{*{20}c}|&|\\
u_v({\bf k},{\bf r},t)&u_c({\bf k},{\bf r},t)\\
|&|\end{array}\right)\in U(2n).
\end{equation}
The orthogonality of conduction and valence band states implies that
 \begin{equation}\label{FPPgc}
G_{{\bf k},{\bf r},t}^\dagger\Xi G_{-{\bf k},{\bf r},t}=0.
\end{equation}
In general, we call a frame $G:\partial\mathcal{T}_{1/2}\to U(2n)$
particle-hole trivial if it can continuously be deformed to
a constant {\it while satisfying \eqref{FPPgc} throughout the
deformation}.  The Chern Simons term in \eqref{FPPFK} requires a
gauge that is built from the valence band part of a particle-hole
trivial frame.

Though the subtlety of the gauge condition makes a general
computation of the invariant difficult, it is
possible to understand the invariant in
the context of specific models.  Consider, a theory based
on a point defect in the $d$ dimensional version of \eqref{jackiwrossi},
\begin{equation}
{\cal H}({\bf k},{\bf r},t) = v \vec\gamma\cdot{\bf k} +
\vec\Gamma\cdot \vec\phi({\bf r},t).
\end{equation}
Here $\vec\Gamma$ and $\vec\gamma$ are $2^d\times 2^d$ Dirac
matrices, and we suppose that for fixed $t$, the $d$ dimensional mass
vector $\vec \phi({\bf r},t)$ has a point topological defect at ${\bf
r}_0(t)$.  If Ref. \onlinecite{teokane10} we argued that adiabatic cycles for such point
defects are classified by a Pontrjagin invariant similar to that
discussed in section \ref{sec:pointmajorana2d}.  This may also be understood in terms of the
rotation of the ``orientation" of the defect.
Near the defect, suppose $\vec\phi({\bf r},t) = O(t)\cdot ({\bf r}-{\bf
r}_0(t))$, where $O(t)$ is a time dependent $O(d)$ rotation.  In the
course of the cycle, the orientation of the topological defect,
characterized by $O(t)$ goes through a cycle.  Since for $d\ge 3$, $\pi_1(O(d)) =
\mathbb{Z}_2$, there are two classes of cycles.  As shown in Ref. \onlinecite{teokane10},
the non trivial cycle, which corresponds to a $2\pi$ rotation changes
the sign of the Majorana fermion wavefunction associated with the
topological defect.  We will argue below that this corresponds to a
change in the local fermion parity in the vicinity of the defect.
For $d=2$, $\pi_1(O(2)) = \mathbb{Z}$.  However, the change in the
sign of the Majorana bound state is given by the parity of the $O(2)$
winding number.  In theories with more bands, it is only this parity
that is topologically robust.

In $d=1$, the single $\Gamma$ matrix in the 2 band model does not allow for
continuous rotations.  Consider instead Kitaev's model\cite{kitaev00}
for a 1D topological superconductor with at time dependent phase,
\begin{equation}
{\cal H}(k,t) = (t \cos k - \mu )\tau_z + \Delta_1(t) \sin k \tau_x +
\Delta_2(t) \sin k \tau_y.
\end{equation}
In this case it is possible to apply the formula \eqref{FPPFK} because on the
boundary $\partial {\cal T}$, which is $k = 0$ or $k=\pi$ the
Hamiltonian is independent of $t$, so that the gauge condition \eqref{FPPgc} is
automatically satisfied.  Moreover, the second term in \eqref{FPPFK} involving
the Chern Simons integral is equal to zero, so that the invariant is
simply the integral of ${\cal F}(x,t)$ over ${\cal T}_{1/2}$.  It is
straightforward to check that this gives $\nu=1$.

\begin{figure}
\epsfxsize=2.5in
\epsfbox{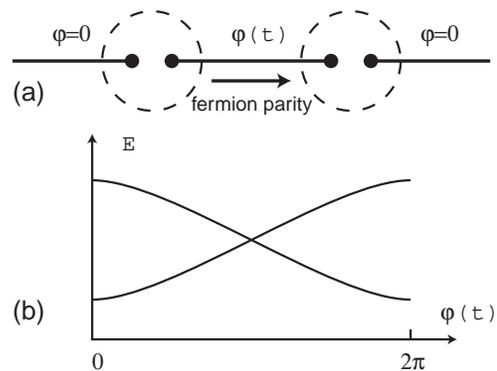}
\caption{A one dimensional fermion parity pump based on a 1D topological
superconductor, which has Majorana states at its ends.  When the phase of the central
superconductor is advanced by $2\pi$ the fermion parity associated with the pairs of
Majorana states inside each circle changes.  Thus fermion parity has been pumped from
one circle to the other.  (b) shows the evolution of the energy levels associated with a
weakly coupled pair of Majorana states as a function of phase.  The level crossing at $\phi=\pi$
is protected by the local conservation of fermion parity.}
\label{fppump}
\end{figure}

In order to see why this corresponds to a pump for fermion parity,
suppose a topological superconductor is broken in two places,
as shown in Fig. \ref{fppump}.  At the ends where the superconductor
is cut there will be Majorana bound states.  The pair of bound states
associated with each cut define two quantum states which differ by
the {\it parity} of the number of electrons.   If the two ends are
weakly coupled by electron tunneling then the pair of states will
split.  Now consider advancing the phase of the central superconductor by
$2\pi$.  As shown in Refs. \onlinecite{kitaev00,fukane09a}, the states interchange as depicted in
Fig. \ref{fppump}.  The level crossing that occurs at $\pi$ phase
difference is protected by the conservation of fermion parity.  Thus,
at the end of the cycle, one unit of fermion parity has been
transmitted from one circled region to the other.

The pumping of fermion parity also applies to adiabatic cycles of point defects
in higher dimensions, and is deeply connected with the Ising non-Abelian
statistics associated with those defects\cite{teokane10}.

\section{Conclusion}
\label{sec:conclusion}

In this paper we developed a unified framework for classifying
topological defects in insulators and superconductors by considering
Bloch/BdG Hamiltonians that vary adiabatically with spatial
(and/or temporal) parameters.  This led to a generalization of the
bulk-boundary correspondence, which identifies protected gapless
fermion excitations with topological invariants characterizing the
defect.  This leads to a number of additional questions to be
addressed in future work.

The generalized bulk-boundary correspondence has the flavor of a
mathematical index theorem, which relates an analytic index that
characterizes the zero modes of a system to a topological index.
It would be interesting to see a more general formulation of this
relation\cite{fukui10a,fukui10b} that applies to the classes without chiral symmetry
that have $\mathbb{Z}_2$ invariants, and goes beyond the adiabatic approximation we used
in this paper.   Though the structure of the gapless modes associated with defects
make it clear that such states are robust in the presence of disorder
and interactions, it would be desirable to have a more general
formulation of the topological invariants characterizing a defect
that can be applied to interacting and/or disordered
problems.

An important lesson we have learned is that topologically protected
modes can occur in a context somewhat more general than simply
boundary modes.  This expands the possibilities for engineering these
states in physical systems.  It is thus an important future direction
to explore the possibilities for heterostructures that realize
topologically protected modes.  The simplest version of this would be
to engineer protected chiral fermion modes using a magnetic
topological insulator.  The perfect electrical transport in such
states could have far reaching implications at both the fundamental
and practical level.   In addition, it is worth considering the
expanded possibilities for realizing Majorana bound states in
superconductor heterostructures, which could have implications for
quantum computing.

Finally, it will be interesting to generalize these topological
considerations to describe inherently correlated states, such as the
Laughlin state.  Could a {\it fractional} quantum Hall edge state
arise as a topological line defect in a 3D system?  Understanding the
topological invariants that would characterize such a defect would
lead to a deeper understanding of topological states of matter.

\acknowledgments

We thank Claudio Chamon, Liang Fu, Takahiro Fukui and Roman Jackiw for helpful discussions.
This work was supported by NSF grant DMR-0906175.

\begin{appendix}

\section{Periodicity in symmetry and dimension}
\label{appendix:periodicities}

In this appendix we will establish the relations (\ref{period1},\ref{period2})
between the $K$-groups in different position-momentum dimensions $(D,d)$ and different symmetry classes $s$.
We will do so by starting with an arbitrary Hamiltonian in $K_{\mathbb{F}}(s;D,d)$ and
then explicitly constructing new Hamiltonians in one higher
position or momentum dimension, which have a
symmetry either added or removed.  The new Hamiltonians will then belong to
$K_{\mathbb{F}}(s+1;D,d+1)$ or $K_{\mathbb{F}}(s-1;D+1,d)$.  The first step
is to identify the mappings and show they preserve the group structure.
This defines group homomorphisms relating the $K$ groups.
The next step is to show they are {\it isomporphisms} by showing that the
maps have an inverse, up to homotopic equivalence.

\subsection{Hamiltonian mappings}
\label{appendix:hamiltonianmaps}

There are two classes of mappings: those that add symmetries and those that
remove symmetries.  These need to be considered separately.

\begin{figure}
\epsfxsize=3.0in
\epsfbox{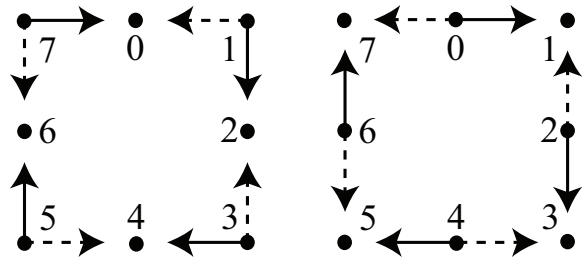}
\caption{Hamiltonian mappings \eqref{ctonc} and \eqref{nctoc} are drawn on
the left and right clocks respectively. Solid (dotted) arrows represents
addition of one momentum (spatial) dimension.}
\label{clockarrow}
\end{figure}

\begin{figure}
\epsfxsize=3.0in
\epsfbox{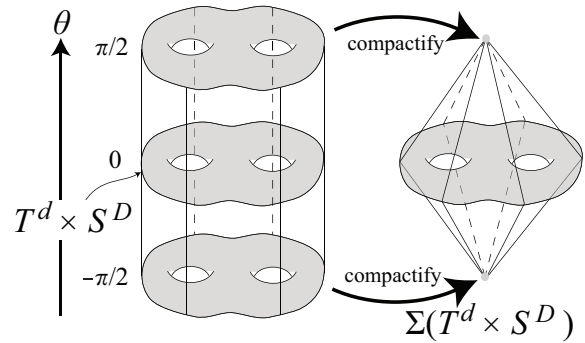}
\caption{Suspension $\Sigma(T^d\times S^D)$. The top and bottom
of the cylinder $\Sigma(T^d\times S^D)\times[-\pi/2,\pi/2]$ are identified to
two points.}
\label{suspension1}
\end{figure}

We consider first
the symmetry removing mappings that send a Hamiltonian ${\cal H}_c$ with chiral symmetry
to a Hamiltonian ${\cal H}_{nc}$ without chiral symmetry.
Suppose $\{{\cal H}_c({\bf k},{\bf r}),\Pi\}=0$, where $\Pi$ is the chiral operator.
Then define
\begin{equation}
{\cal H}_{nc}({\bf k},{\bf r},\theta) = \cos\theta {\cal H}_c({\bf k},{\bf r})
+ \sin\theta \Pi
\label{ctonc}
\end{equation}
for $-\pi/2 \leq \theta \leq \pi/2$.  This has the property that at $\theta = \pm
\pi/2$ the new Hamiltonian is $\pm\Pi$, independent of ${\bf k}$ and ${\bf r}$.
Thus, at each of these points we may consider the base space $T^d \times S^D$
defined by ${\bf k}$ and ${\bf
r}$ to be contracted to a point.  The new Hamiltonian is then defined on
the {\it suspension} $\Sigma(T^d\times S^D)$ of the original base space (see fig.\ref{suspension1}).  If we
treat the original base space as a $d+D$ dimensional sphere, then the suspension
is a $d+D+1$ dimensional sphere.

Without loss of generality we assume ${\cal H}_c$ is flattened, so that
${\cal H}_c^2 = 1$.  Since $\{{\cal H}_c,\Pi\}=0$ it follows that
${\cal H}_{nc}^2 = 1$ as well.
The second term in \eqref{ctonc} violates the chiral symmetry.  Thus, if ${\cal H}_c$
belongs to the complex class AIII (with no anti unitary symmetries), then
${\cal H}_{nc}$ belongs to class A.  Eq.\eqref{ctonc} thus provides a mapping
from class AIII to class A.

For the real classes, which have anti unitary
symmetries, the second term will violate either particle-hole symmetry or
time reversal symmetry, depending on whether $\theta$ is a momentum or position
type variable (odd or even under $\Theta$ and $\Xi$).  This will lead to a new
non chiral symmetry class related to the original class by either
a clockwise or counter clockwise
turn on the symmetry clock (fig.\ref{clockarrow}).  To determine which it is, note that
if we require $[\Theta,\Xi]=0$ then $(\Theta\Xi)^2 = \Theta^2 \Xi^2 =
(-1)^{(s-1)/2}$.  The unitary chiral symmetry operator (satisfying
$\Pi^2=1$) can then be written
\begin{equation}
\Pi = i^{(s-1)/2} \Theta \Xi.
\end{equation}
It follows that if $\theta$ is momentum like, then time reversal symmetry is
violated when $s = 1\ {\rm mod}\ 4$, while particle hole is violated when $s =
3\ {\rm mod}\ 4$.  This corresponds to corresponds to a clockwise rotation on
the symmetry clock, $s \rightarrow s+1$.  If $\theta$
is position like then $s \rightarrow s-1$.

We next build a chiral Hamiltonian from a non chiral one by adding a symmetry.
This is accomplished
by doubling the number of bands in a manner similar to the doubling employed in
the Bogoliubov de Gennes description of a superconductor.  We thus write
\begin{equation}
{\cal H}_c({\bf k},{\bf r},\theta) =  \cos\theta {\cal H}_{nc}({\bf k},{\bf r})
\otimes \tau_z + \sin \theta \openone\otimes\tau_a,
\label{nctoc}
\end{equation}
where $a = x$ or $y$.  Here $\vec\tau$ are Pauli matrices that act on the
doubled degree of freedom.  As in \eqref{ctonc}, \eqref{nctoc} gives a new
Hamiltonian defined on a base space that is the suspension of the original base
space.  If ${\cal H}_{nc}^2 = 1$ it follows that ${\cal H}_c^2=1$, so the energy
gap is preserved.
It is also clear that the new Hamiltonian has a chiral symmetry because it
anticommutes with $\Pi = i \tau_z \tau_a$.  Thus, if ${\cal H}_{nc}$ is in
class A, then ${\cal H}_c$ is in class AIII.

For the real symmetry classes $a=x$ or $y$ must be chosen so
that the second term in \eqref{nctoc} preserves the original anti unitary
symmetry of ${\cal H}_{nc}$.  This depends on the original anti unitary
symmetry and whether $\theta$ is chosen to be
a momentum or a position variable.  For example, if ${\cal H}_{nc}$ has time
reversal symmetry, $\Theta$, and $\theta$ is a momentum (position) variable, then
we require $a=y$ ($a=x$).
In this case, ${\cal H}_c$ has the additional particle-hole symmetry
$\Xi = \tau_x \Theta$ ($\Xi = i\tau_y \Theta$) that satisfies $\Xi^2 =
\Theta^2$ ($\Xi^2 = -\Theta^2$).  A similar analysis when ${\cal H}_{nc}$ has
particle hole symmetry allows us to conclude
that the symmetry class of ${\cal H}_c$ is given
by a clockwise rotation on the symmetry clock, $s \rightarrow s+1$,
when $\theta$ is a momentum variable.  When $\theta$ is a position
variable, $s \rightarrow s-1$ gives a counter clockwise
rotation.

Equations \eqref{ctonc} and \eqref{nctoc} map a Hamiltonian into a new Hamiltonian
in a different dimension and different symmetry class.  It is clear that two
Hamiltonians that are topologically equivalent will be mapped to topologically
equivalent Hamiltonians, since the mapping can be done continuously on a smooth
interpolation between the original Hamiltonians.  Thus,
(\ref{ctonc},\ref{nctoc}) define a mapping between equivalence classes of
Hamiltonians.  Moreover, since the direct sum of two Hamiltonians is mapped to
the direct sum of the new Hamiltonians, the group property of the equivalence
classes is preserved. (\ref{ctonc},\ref{nctoc}) thus define a $K$-group
homomorphism,
\begin{eqnarray}
K_{\mathbb{F}}(s;D,d)&\longrightarrow&K_{\mathbb{F}}(s+1;D,d+1)\label{arrowm},\\
K_{\mathbb{F}}(s;D,d)&\longrightarrow&K_{\mathbb{F}}(s-1;D+1,d)\label{arrowp},
\end{eqnarray}
for $\mathbb{F}=\mathbb{R},\mathbb{C}$.

\subsection{Invertibility}
\label{appendix:invertivility}

In order to establish that (\ref{arrowm},\ref{arrowp}) are isomorphisms we need
to show that there exists an inverse.  This is {\it not} true of the
Hamiltonian mappings.  A general Hamiltonian cannot
be built from a lower dimensional Hamiltonian using (\ref{ctonc},\ref{nctoc}).
However, we will argue that it is possible to continuously deform any
Hamiltonian into the form given by \eqref{ctonc} or \eqref{nctoc}.  Thus,
the mappings between equivalence classes have an inverse.  To show this we will
use a mathematical method borrowed from Morse theory\cite{JMilnor}.

Without loss of generality we again consider {\em flattened} Hamiltonians having
equal number of conduction and valence bands with energies $\pm 1$.
Consider ${\cal H}({\bf k},{\bf r},\theta)$, where $\theta \in [-\pi/2,\pi/2]$
is either a position
or momentum variable and ${\cal H}$ is independent of ${\bf k}$ and ${\bf r}$
at $\theta = \pm\pi/2$.  We wish to show that ${\cal H}({\bf k},{\bf r},\theta)$
can be continuously deformed into the form \eqref{ctonc} or \eqref{nctoc}.
To do so we define an artificial ``action"
\begin{equation}
\label{Haction}
S[\mathcal{H}({\bf k},{\bf r},\theta)]=\int d\theta d^d{\bf k}d^D{\bf
r}\mbox{Tr}\left(\partial_\theta\mathcal{H}\partial_\theta\mathcal{H}\right).
\end{equation}
$S$ can be interpreted as a ``height" function in the space
of gapped symmetry preserving Hamiltonians.
Given any Hamiltonian there is always a downhill
direction.  These downhill vectors can then be integrated into a deformation
trajectory. Since the action is
positive definite, it is bounded below. The deformation trajectory must end at
a Hamiltonian that locally minimizes the action.

Under the flatness constraint $\mathcal{H}^2=1$, minimal Hamiltonians satisfy
the Euler-Lagrange equation
\begin{equation}
\label{HEL}
\partial^2_\theta\mathcal{H}+\mathcal{H}=0.
\end{equation}
The solutions must be a linear combination of $\sin\theta$ and $\cos\theta$.  The
coefficient of $\sin\theta$ must be constant because the base space is
compactified to points at $\theta=\pm\pi/2$.  A minimal Hamiltonian thus has the
form
\begin{equation}
\mathcal{H}({\bf k},{\bf
r},\theta)=\cos\theta\mathcal{H}_1({\bf k},{\bf
r})+\sin\theta\mathcal{H}_0\label{Asol}.
\end{equation}
The constraint
$\mathcal{H}({\bf k},{\bf r},\theta)^2=1$ requires
\begin{equation}
\mathcal{H}_0^2=\mathcal{H}_1({\bf k},{\bf
r})^2=1,\quad\{\mathcal{H}_0,\mathcal{H}_1({\bf k},{\bf
r})\}=0.
\label{pAsol}
\end{equation}

If $\mathcal{H}({\bf k},{\bf r},\theta)$ is non-chiral, then eq.\eqref{Asol} is
already in the form of eq.\eqref{ctonc} with $\Pi=\mathcal{H}_0$
and $\mathcal{H}_c({\bf k},{\bf r}) = \mathcal{H}_1({\bf k},{\bf r})$.
$\mathcal{H}_1$ automatically has
chiral symmetry due to Eq.\eqref{pAsol}.  This shows that \eqref{arrowm} and
\eqref{arrowp} are invertible when $s$ is odd.

If $\mathcal{H}({\bf k},{\bf r},\theta)$ is chiral, then both $\mathcal{H}_0$
and $\mathcal{H}_1({\bf k},{\bf r})$ anticommute with the chiral symmetry
operator $\Pi$. Rename $\mathcal{H}_0=\tau_a$ and $\Pi=i\tau_z \tau_a$, where
$a = x$ ($a=y$) when $\theta$ is a position (momentum) variable.  It follows
that
$\{\mathcal{H}_1,\tau_x\}=\{\mathcal{H}_1,\tau_y\}=0$, so we can write
\begin{equation}
\mathcal{H}_1({\bf k},{\bf r})=h({\bf k},{\bf r})\otimes\tau_z.
\end{equation}
Eq.\eqref{Asol} thus takes the form of Eq.\eqref{nctoc} with
${\cal H}_{nc} = h$.  Since $\tau_z$ anti-commutes with either $\Theta$ or
$\Xi$, $h({\bf k},{\bf r})$ carries
exactly one anti-unitary symmetry and is therefore non-chiral.
This shows that (\ref{arrowm}) and
(\ref{arrowp}) are invertible when $s$ is even.

\section{Representative Hamiltonians, and classification by winding numbers}
\label{appendix:representatives}

In this appendix we construct representative Hamiltonians for each
of the symmetry classes that are built as linear combinations of
Clifford algebra generators that can be represented as anticommuting Dirac matrices.
This allows us to relate the integer topological invariants, corresponding to the
$\mathbb{Z}$ and $2\mathbb{Z}$ entries in Table I, to the winding degree in maps between
spheres. Similar construction for defectless bulk Hamiltonians
can be found in Ref.[\onlinecite{ryu10}] by Ryu, {\it et.al}.
In general, Hamiltonians do not have this specific form.
However, since each topological class of Hamiltonians  includes
representatives of this form, it is always possible to smoothly
deform ${\cal H}({\bf k},{\bf r})$ into this form.

The simplest example of this approach is the familiar case of a two
dimensional Hamiltonian with no symmetries (class $A$).  A
topologically non trivial Hamiltonian can be represented as a
$2\times 2$ matrix that can be expressed in terms of Pauli matrices
as ${\cal H}({\bf k}) = {\bf h}({\bf k})\cdot\vec\sigma$.  The
Hamiltonian can then be associated with a unit vector $\hat {\bf d}({\bf
k}) = {\bf h}({\bf k})/|{\bf h}({\bf k})| \in S^2$.  It is then
well known that the Chern number characterizing ${\cal H}({\bf k})$
in two dimensions is related to the {\em degree}, or winding number, of the mapping
from ${\bf k}$ to $S^2$.  This approach also applies to higher Chern
numbers characterizing Hamiltonians in even dimensions $d=2n$.  In
this case, a Hamiltonian that is a combination of $2n+1$ $2^n\times 2^n$
Dirac matrices, and can be associated with a unit vector $\hat{\bf
d} \in S^{2n}$.

For the complex chiral class AIII, the $U(n)$ winding number
characterizing a family of Hamiltonians can similarly be expressed as
a winding number on spheres.  For example, in $d=1$, a chiral
Hamiltonian can be written ${\cal H}(k) = h_x(k) \sigma_x +
h_y(k)\sigma_y$ (so $\{{\cal H},\sigma_z\}=0$), and is characterized
by $\hat{\bf d}(k) \in S^1$.  The integer
topological invariant can then be expressed by the winding number of
$\hat{\bf d}(k)$.  Similar considerations apply to the integer
invariants for chiral Hamiltonians in higher odd dimensions.

For the real symmmetry classes we introduce
``position type" Dirac matrices $\Gamma_\mu$ and ``momentum type"
Dirac matrices $\gamma_i$.  These satisfy
$\{\Gamma_\mu,\Gamma_\nu\}=2\delta_{\mu\nu}$,
$\{\gamma_i,\gamma_j\}=2\delta_{ij}$, $\{\Gamma_\mu,\gamma_j\}=0$
and are distinguished by their symmmetry under anti unitary
symmetries. If there is time reversal symmetry we require
\begin{equation}
[\Gamma_\mu,\Theta] = \{\gamma_i,\Theta\} = 0,
\end{equation}
while with particle-hole symmetry,
\begin{equation}
\{\Gamma_\mu,\Xi\} = [\gamma_i,\Xi] = 0.
\end{equation}

For a Hamiltonian that is a combination of $p$ momentum like matrices
$\gamma_{1,\ldots,p}$ and $q+1$ position like matrices
$\Gamma_{0,\ldots,q}$,
\begin{equation}
\mathcal{H}({\bf k},{\bf r})={\bf R}({\bf k},{\bf
r})\cdot\vec\Gamma+{\bf K}({\bf k},{\bf
r})\cdot\vec\gamma,
\label{Hmodel}
\end{equation}
the coefficients must satisfy the involution
\begin{eqnarray}
{\bf R}(-{\bf k},{\bf r})&=&{\bf R}({\bf k},{\bf r}),\label{Diracinvol1}\\
{\bf K}(-{\bf k},{\bf r})&=&-{\bf K}({\bf k},{\bf r})\label{Diracinvol2}.
\end{eqnarray}
This can be characterized by a unit vector
\begin{equation}
\hat{\bf d}({\bf k},{\bf r}) = {\frac{({\bf K},{\bf R})}{\sqrt{|{\bf
K}|^2 + |{\bf R}|^2}}} \in S^{p+q},
\end{equation}
where $S^{p+q}$ is a $(p+q)$-sphere in which $p$ of the dimensions
are odd under the involution (\ref{Diracinvol2}).

The symmetry class $s$ of ${\cal H}({\bf k},{\bf r})$ is related to the
indices $(p,q)$ characterizing the numbers of Dirac matrices by
\begin{equation}
p-q = s \ {\rm mod} \ 8.
\label{p-q}
\end{equation}
To see this, start with a Hamiltonian ${\cal H}_0 = R_0({\bf k},{\bf r})
\Gamma_0$ that involves a single $1\times 1$ position like
``Dirac matrix" $\Gamma_0 = \openone$, so $(p,q)=(0,0)$.  This clearly has
time reversal symmetry, with $\Theta = K$, and corresponds to class AI
with $s=0$.
Next, generate Hamiltonians ${\cal H}_s$ with different
symmetries $s$ by using the Hamiltonian mappings introduced in appendix \ref{appendix:hamiltonianmaps}.
Both the mappings \eqref{ctonc} and \eqref{nctoc} define a new
Clifford algebra with one extra generator that is either position or momentum type.
The mappings that correspond to clockwise rotations on the symmetry clock
($s\rightarrow s+1$) introduce an additional position like generator ($p
\rightarrow p+1$), while the mappings that correspond to counterclockwise
rotations ($s\rightarrow s-1$) introduce an additional momentum like
generator ($q \rightarrow q+1$).  Eq.\eqref{p-q} follows because
this procedure can be repeated to generate Hamiltonians with any indices $(p,q)$.
Some examples are listed in table \ref{exDiracMatrix}
\begin{table}
\centering
\begin{ruledtabular}
\begin{tabular}{cc|ccccc|ccc}
\multicolumn{2}{c|}{Classes} & \multicolumn{5}{c|}{Dirac matrices} & \multicolumn{3}{c}{Symmetry operators} \\
s & AZ & $\Gamma_0$ & $\vec\gamma$ &  &  &  & $\Theta$ & $\Xi$ & $\Pi$ \\\hline
0 & AI & \openone &  &  &  &  & $K$ &  &  \\
1 & BDI & $\sigma_z$ & $\sigma_y$ &  &  &  & $K$ & $\sigma_xK$ & $\sigma_x$ \\
2 & D & $\sigma_z$ & $\sigma_y$ & $\sigma_x$ &  &  &  & $\sigma_xK$ &  \\
3 & DIII & $\tau_z\sigma_z$ & $\tau_z\sigma_y$ & $\tau_z\sigma_x$ & $\tau_x$ &  & $i\tau_y\sigma_xK$ & $\sigma_xK$ & $\tau_y$ \\
4 & AII & $\tau_z\sigma_z$ & $\tau_z\sigma_y$ & $\tau_z\sigma_x$ & $\tau_x$ & $\tau_y$ & $i\tau_y\sigma_xK$ &  &  \\
\end{tabular}
\end{ruledtabular}
\caption{Examples of Dirac matrices for $(p,q)=(s,0)$. }
\label{exDiracMatrix}
\end{table}

The integer topological invariants in Table \ref{tab:periodic} (which occur when
$s-\delta$ is even) can be related to the
winding degree of the maps $\hat {\bf d}: S^{D+d} \rightarrow
S^{p+q}$.  This can be non zero when the spheres have the same
total dimensions. In light of \eqref{p-q}, $(p,q)$ can always
be chosen so that $d+D=p+q$.
The anti unitary symmetries impose constraints on the possible values
of these winding numbers, which depend on the relation between
$\delta = d-D$ and $s=p-q$.

The involutions on $S^{d+D}$ and $S^{p+q}$ have opposite orientations when
$\delta-s\equiv2$ or 6 mod 8, and therefore an involution preserving map
$S^{d+D}\to S^{p+q}$ can have non-zero winding degree only when
$\delta-s\equiv0$ or 4 mod 8.  Symmetry gives a further constraint on the latter
case. Consider a sphere map $S^{2}_{\theta,\phi}\to
S^{2}_{\vartheta,\varphi}$, where the involutions on the spheres send
$(\theta,\phi)\mapsto(\theta,\phi+\pi)$ and
$(\vartheta,\varphi)\mapsto(\vartheta,\varphi)$. In order for
$\varphi(\theta,\phi)=\varphi(\theta,\phi+\pi)$, the winding number must be
even. Together, these show
\begin{equation}\deg\in\left\{\begin{array}{*{20}c}\mathbb{Z},&\mbox{for
$\delta-s\equiv0$ (mod 8)}\\2\mathbb{Z},&\mbox{for $\delta-s\equiv4$ (mod
8)}\\0,&\mbox{otherwise}\hfill\end{array}\right.\end{equation}
This gives a topological understanding of the $\mathbb{Z}$'s and
$2\mathbb{Z}$'s on the periodic table in terms of winding number, which can be
identified with the more general analytic invariants, namely Chern numbers for
non-chiral classes
\begin{equation}n=\frac{1}{\left(\frac{d+D}{2}\right)!}\left(\frac{i}{2\pi}\right)^{\frac{d+D}{2}}
\int_{T^d\times S^D}\mbox{Tr}\left(\mathcal{F}^{\frac{d+D}{2}}\right)\end{equation}
 and winding numbers of the chiral flipping operator
$q({\bf k},{\bf r})$ for chiral ones (See Eqs. \ref{hchiral} and
\ref{nchiral})).
\begin{equation}n=\frac{\left(\frac{d+D-1}{2}\right)!}{(d+D)!(2\pi i)^{\frac{d+D+1}{2}}}
\int_{T^d\times S^D}\mbox{Tr}\left((qdq^\dagger)^{d+D}\right)\end{equation}

The $\mathbb{Z}_2$'s on the periodic table are not directly
characterized by winding degree, but rather through dimensional
reduction. Given a Hamiltonian $\mathcal{H}({\bf k},k_1,k_2,{\bf r})$
with $s\equiv\delta$ mod 8, its winding degree mod 2 determines the
$\mathbb{Z}_2$-classification of its equatorial offspring
$\mathcal{H}_{k_2=0}({\bf k},k_1,{\bf r})$ and
$\mathcal{H}_{k_{1,2}=0}({\bf k},{\bf r})$. For example, topological
insulators in two and three dimensions are equatorial restrictions of
a four dimensional model $\hat{\bf d}:S^{4}\xrightarrow{=}S^{4}$
with unit winding number. Around the north pole, the Hamiltonian has
the form
\begin{equation}
\mathcal{H}({\bf k},k_4)=(m+\varepsilon
k^2)\mu_1+{\bf k}\cdot\mu_3\vec\sigma+k_4\mu_2
\end{equation}
and on the equator $k_4=0$, this gives a three dimensional Dirac theory of
mass $m$ around ${\bf k=0}$ that locally describes 3D topological
insulators $\mbox{Bi}_2\mbox{Se}_3$ and Bi$_2$Te$_3$ around $\Gamma$.

\section{Zero Modes in the Harmonic Oscillator Model}
\label{appendix:ZM}

We present exact solvable soliton states of Dirac-type defect
Hamiltonians. These include zero modes at a point defect of a
Hamiltonian in the chiral class AIII, and chiral modes along a line
defect of a Hamiltonian in the non-chiral class A. We establish the
connection between the two kinds of boundary modes through
the Hamiltonian mapping \eqref{ctonc}.

A non-trivial chiral Hamiltonian isotropic around a point defect at
${\bf r}=0$ is a Dirac operator
\begin{equation}\mathcal{H}=-i\vec\gamma\cdot\nabla+{\bf
r}\cdot\vec\Gamma\label{app-chd}\end{equation} where the chiral
operator is $\Pi=i^d\prod_{j=1}^d\gamma_j\Gamma_j$, and its adiabatic
limit $e^{-i{\bf k}\cdot{\bf r}}\mathcal{H}e^{i{\bf k}\cdot{\bf
r}}={\bf k}\cdot\vec\gamma+{\bf r}\cdot\vec\Gamma$ has unit winding
degree on $S^{2d-1}=\{({\bf k},{\bf
r}):k^2+r^2=1\}$.\begin{equation}\mathcal{H}^2=-\nabla^2+r^2-i\vec\gamma\cdot\vec\Gamma\end{equation}
and the spectrum is determined by the quantum numbers $n_j\geq0$ of
the harmonic oscillator and the parities $\xi_j$ of the mutually
commuting matrices $i\gamma_j\Gamma_j$, for $j=1,\ldots,d$.
\begin{equation}
\mathcal{E}^2=\sum_{j=1}^d 2n_j+1 -\xi_j
\end{equation}
The unique zero energy state $|\Psi_0\rangle$, indexed by $n_j=0$ and
$\xi_j=1$, has positive chirality $\Pi=+1$, and is exponentially
localized at the point defect as $\Psi_0({\bf r})\propto
e^{-\frac{1}{2}r^2}$.

Next we consider a non-chiral Hamiltonian isotropic along a line
defect.
\begin{equation}
\mathcal{H}(k_\|)=k_\|\Pi-i\vec\gamma\cdot\nabla+{\bf
r}\cdot\vec\Gamma\label{app-nhd}
\end{equation} where $k_\|$ is
parallel to the defect line, ${\bf r}$ and $\nabla$ are normal
position and derivative. Its adiabatic limit $e^{-i{\bf k}\cdot{\bf
r}}\mathcal{H}(k_\|)e^{i{\bf k}\cdot{\bf r}}=k_\|\Pi+{\bf
k}\cdot\vec\gamma+{\bf r}\cdot\vec\Gamma$ is related to that of
(\ref{app-chd}) by (1,1)-periodicity, and has unit winding degree on
$S^{2d}=\{(k_\|,{\bf k},{\bf r}):k_\|^2+|{\bf k}|^2+|{\bf
r}|^2=1\}$. The zero mode $|\Psi_0\rangle$ of (\ref{app-chd}) gives
rise to a positive chiral mode,
$\mathcal{H}(k_\|)|\Psi_0\rangle=k_\|\Pi|\Psi_0\rangle=+k_\||\Psi_0\rangle$.

\begin{figure}\label{engplot}
 \centerline{ \epsfig{figure=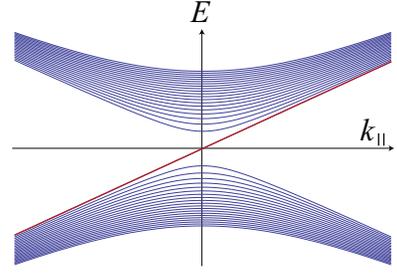,width=2.0in}}
 \caption{Energy spectrum of Hamiltonian (\ref{app-nhd}). The zero mode
 $|\Psi_0\rangle$ of positive chirality at $k_\|=0$ corresponds the
 chiral mode that generates the mid-gap $k_\|$-linear energy spectrum.}
\end{figure}

The two examples verified bulk-boundary correspondence through
identifying analytic information of the defect-bound solitons and the
topology of slowly spatial modulated theories far away from the
defect. The single zero mode of (\ref{app-chd}) and spectral flow of
(\ref{app-nhd}) are equated to unit winding degree of an adiabatic
limit. In general, bulk-boundary correspondence is mathematically
summarized by index theorems that associate certain analytic and
topological indices of Hamiltonians.\cite{jackiwrebbi,jackiwrossi,weinberg,volovik03,fukui10a,fukui10b}

\section{Invariant for Point Defects in Class D and BDI}
\label{appendix:pointdefectclassd}

We follow the derivation given in Ref.[\onlinecite{teokane10}], which was based on Qi, Hughes and Zhang's
formulation of the topological invariant characterizing a three dimensional
topological insulator\cite{qihugheszhang08}.  For a point defect in $d$ dimensions, the Hamiltonian
${\cal H}({\bf k},{\bf r})$ depends on $d$ momentum variables and $d-1$
position variables.  We introduce a one parameter deformation $\widetilde{\cal H}(\lambda,{\bf
k},{\bf r})$ that connects $\widetilde{\cal H}({\bf k},{\bf r})$ at $\lambda =0$ to a
constant Hamiltonian at $\lambda = 1$, while breaking particle-hole symmetry.
The particle-hole symmetry can be restored by including a mirror image
$\tilde {\cal H}(\lambda,{\bf k},{\bf r}) = - \Xi {\cal H}(-\lambda,{\bf
k},{\bf r})\Xi^{-1}$ for $-1<\lambda<0$.  For $\lambda = \pm $,
$({\bf k},{\bf r})$ can be replaced by a single point, so the $2d$ parameter
space $(\lambda,{\bf k},{\bf r})$ is the suspension $\Sigma(T^d\times S^{d-1})$
of the original space.  The Hamiltonian defined on this space is characterized
by its $d$'th Chern character
\begin{equation}
\nu = {\frac{1}{d!}} \left({\frac{i}{2\pi}}\right)^d \int_{\Sigma(T^d\times S^{d-1})} {\rm
Tr}[ {\cal F}^d ].
\end{equation}
Due to particle-hole symmetry, the contributions from the two hemispheres
$\lambda >0$, $\lambda<0$ are equal.  Using the fact that the integrand is the
derivative of the Chern Simons form,
${\rm Tr}[{\cal F}^d] = d{\cal Q}_{2d-1}$, we can therefore write
\begin{equation}\label{CSD}
\nu = \frac{2}{d!}\left(\frac{i}{2\pi}\right)^d \int_{T^d\times S^{d-1}} {\cal Q}_{2d-1}
\end{equation}
As was the case in Refs.[\onlinecite{teokane10,qihugheszhang08}], $\nu$ can be different for different deformations ${\cal
H}(\lambda,{\bf k},{\bf r})$.  However, particle-hole symmetry requires the
difference is an even integer.  Thus, the parity of \eqref{CSD} defines the
$\mathbb{Z}_2$ invariant.

The Chern Simons form $\mathcal{Q}_{2d-1}$ can be expressed in terms of the
connection $\mathcal{A}$ via the general formula
\begin{equation}\label{defCS}
\mathcal{Q}_{2d-1}=d\int_0^1dt\mbox{Tr}\left[\mathcal{A}(td\mathcal{A}+t^2\mathcal{A}^2)^{d-1}
\right]
\end{equation}

In the addition of time reversal symmetry $\Theta^2=1$ or equivalently a chiral
symmetry $\Pi=\Theta\Pi=\tau_z$, a valence frame of the BDI Hamiltonian \eqref{hchiral}
can be chosen to be
\begin{equation}
u({\bf k},{\bf
r})=\frac{1}{\sqrt{2}}\left(\begin{array}{*{20}c}q({\bf k},{\bf
r})\\-\openone\end{array}\right)
\end{equation}
where $q$ is unitary and
$\openone$ is the identity matrix. This corresponds the Berry connection
$\mathcal{A}=u^\dagger du=\frac{1}{2}q^\dagger dq$ and Chern-Simons form
\begin{eqnarray}
Q_{2d-1}&=&\frac{d}{2}\int_0^1dt\left(\frac{t}{2}\left(\frac{t}{2}-1\right)\right)^{d-1}\mbox{Tr}\left[(q^\dagger
dq)^{2d-1}\right]\nonumber\\&=&\frac{(-1)^d}{2}\frac{d!(d-1)!}{(2d-1)!}\mbox{Tr}\left[(q^\dagger
dq)^{2d-1}\right]\label{CS=wn/2}
\end{eqnarray}
This equates the winding number of $q$ \eqref{nchiral}
to the Chern Simons invariant \eqref{muchernsimons}.

\section{Invariant for line defects in class AII and point defects in DIII}
\label{appendix:defectsaiidiii}

We formulate a topological invariant that characterizes
line defects in class AII in all dimensions that is analogous to the
integral formula invariant characterizing the quantum spin Hall insulator
introduced in Ref. \onlinecite{fukane06}.  This can be applied to
weak topological insulators in three dimensions with dislocation around
a line defect. The invariant can be indirectly applied to strong
topological insulators through decomposition into strong and weak
components. As a consequence of the Hamiltonian mapping \eqref{ctonc}
that identifies $(s=4,\delta=2)$ and $(s=3,\delta=1)$, this gives a new topological
invariant that classified point defects in class DIII in all dimensions.

\subsection{Line defects in class AII}
\label{appendix:defectsaii}

\begin{figure}
\epsfxsize=3.0in
\epsfbox{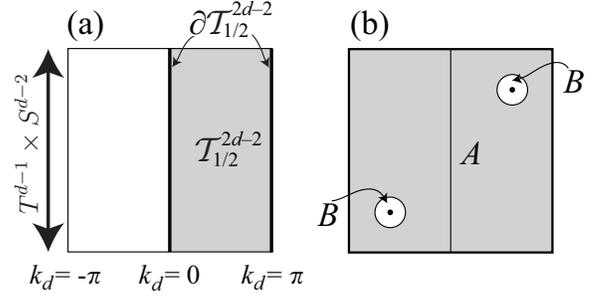}
\caption{
 (a) Schematic of the base space $\mathcal{T}^{2d-2}=T^d\times S^{d-2}$, split into two halves.
 (b) Division of $\mathcal{T}^{2d-2}$ into patches $A$ and $B$,
 each closed under TR and has an individual valence frame $|u_m^{A/B}\rangle$
 that satisfies the gauge condition (\ref{FKgc}).}
\label{FK31}
\end{figure}

The base space manifold is $\mathcal{T}^{2d-2}=T^d\times S^{d-2}$,
where $T^d$ is the Brillouin zone and $S^{d-2}\times\mathbb{R}$ is a
cylindrical neighborhood that wraps around the line defect in real space.
 Divide the base space into two pieces, $\mathcal{T}^{2d-2}_{1/2}$
 and its time reversal counterpart (see fig.\ref{FK31}(a)). We will show
 the $\mathbb{Z}_2$-invariant
 \begin{equation}\nu=\frac{i^{d-1}}{(d-1)!(2\pi)^{d-1}}
 \left[\int_{\mathcal{T}^{2d-2}_{1/2}}\mbox{Tr}(\mathcal{F}^{d-1}) -\oint_{\partial\mathcal{T}^{2d-2}_{1/2}}\mathcal{Q}_{2d-3}\right]
 \label{FKf}\end{equation}
 topologically classifies line defects in AII, where the Chern-Simons
 form, defined by \eqref{defCS}, is generated by the Berry connection
$\mathcal{A}_{mn}=\langle u_m({\bf k},{\bf r})|du_n({\bf k},{\bf r})\rangle$,
and the valence frame $u_m({\bf k},{\bf r})$ satisfies
the gauge condition \begin{equation}\label{FKgc}w_{mn}({\bf k},{\bf r})
 =\langle u_m({\bf k},{\bf r})|\Theta u_n(-{\bf k},{\bf r})\rangle=
 \mbox{constant}\end{equation}
on the boundary $({\bf k},{\bf r})\in\partial\mathcal{T}^{2d-2}_{1/2}$.

The non-triviality of the $\mathbb{Z}_2$-invariant is a topological
obstruction to choosing a global continuous valence frame
$|u_m({\bf k},{\bf r})\rangle$ that satisfies the gauge condition
(\ref{FKgc}) on the whole base space $\mathcal{T}^{2d-2}$. If there
is no topological obstruction from the bulk\cite{ldtopobst}, the gauge
condition forces the valence frame to be singular at two points, depicted
in fig.\ref{FK31}(b), related to each other by time reversal. One removes
the singularity by picking another valence frame locally defined on two
small balls enclosing the two singular points, denoted by $B$ in
fig.\ref{FK31}(b). We therefore have two valence frames
$|u_m^{A/B}({\bf k},{\bf r})\rangle$ defined on two patches of the base space,
$A=\mathcal{T}^{2d-2}\backslash B$ and B, each obeying the gauge
condition (\ref{FKgc}).

The wavefunctions on the two patches translate into each other
through transition function \begin{equation}t^{AB}_{mn}({\bf k},{\bf r})
=\langle u_m^A({\bf k},{\bf r})|u_n^B({\bf k},{\bf r})\rangle\in U(k)
\end{equation} on the boundary $\partial B\approx S^{2d-3}\cup S^{2d-3}$.
The function behavior on the two disjoint $(2d-3)$-spheres is related
by time reversal. The topology is characterized by the winding of
$t^{AB}:S^{2d-3}\to U(k)$ on one of the spheres.
\begin{eqnarray}\nu&=&\frac{(d-2)!}{(2d-3)!(2\pi i)^{d-1}}\oint_{S^{2d-3}}\mbox{Tr}\left[\left(t^{AB}d(t^{AB})^\dagger\right)^{2d-3}\right]
\quad\quad\label{windingFK}\\&=&\frac{(-1)^d}{(d-1)!(2\pi i)^{d-1}}\oint_{S^{2d-3}}\left(\mathcal{Q}_{2d-3}^A-\mathcal{Q}_{2d-3}^B\right)
\label{FKtf}\end{eqnarray} The two integrals can be evaluated
separately. Since $d\mathcal{Q}_{2d-3}=\mbox{Tr}(\mathcal{F}^{d-1})$,
Stokes' theorem tells us
\begin{eqnarray}\int_{A\cap \mathcal{T}^{2d-2}_{1/2}}\mbox{Tr}(\mathcal{F}^{d-1})
&=&\left(\oint_{\partial\mathcal{T}^{2d-2}_{1/2}}
-\oint_{S^{2d-3}}\right)\mathcal{Q}_{2d-3}^A\nonumber\\
\int_{B\cap \mathcal{T}^{2d-2}_{1/2}}\mbox{Tr}(\mathcal{F}^{d-1})
&=&\oint_{S^{2d-3}}\mathcal{Q}_{2d-3}^B\nonumber\end{eqnarray}
Combining these into eq.(\ref{FKtf}) identifies the $\mathbb{Z}_2$-invariant
(\ref{FKf}) with the winding number of the transition function.

The curvature term in (\ref{FKf}) is gauge invariant. Any gauge
transformation on the boundary $\partial\mathcal{T}^{2d-2}_{1/2}$
respecting the gauge condition (\ref{FKgc}) has even winding number
and would alter the Chern-Simons integral by an even integer.
The gauge condition is therefore essential to make the formula non-vacuous.

\subsubsection{Spin Chern number}
\label{appendix:spinchernnumber}

A quantum spin Hall insulator is characterized by its spin Chern number
$n_\sigma=(n_\uparrow-n_\downarrow)/2$. We generalize this to time reversal
invariant line defects of all dimensions by equating it with Eq.(\ref{FKf}).
This applies in particular to a model we considered for a linear Josephson
junction in section \ref{sec:linehelicalmajorana}.

A spin operator $S$ is a unitary operator, square to unity, commutes with
the Hamiltonian, and anticommutes with the time reversal operator.
The valence spin frame \begin{equation}|u_m^\downarrow({\bf k},{\bf r})\rangle
=\Theta|u_m^\uparrow(-{\bf k},{\bf r})\rangle
\end{equation} automatically satisfies the time reversal
gauge constraint (\ref{FKgc}). It is straightforward to check that the
curvature and Chern-Simons form can be split as direct sums according
to spins. \begin{eqnarray}\mathcal{F}({\bf k},{\bf r})&=&\mathcal{F}^\uparrow({\bf k},{\bf r})\oplus
\mathcal{F}^\uparrow(-{\bf k},{\bf r})^\ast\\
\mathcal{Q}_{2d-3}({\bf k},{\bf r})&=&\mathcal{Q}_{2d-3}^\uparrow({\bf k},{\bf r})\oplus
\mathcal{Q}_{2d-3}^\uparrow(-{\bf k},{\bf r})^\ast\end{eqnarray}
Again assuming that there is no lower dimensional ``weak" topology,
the $\uparrow$-frame can be defined everywhere on $\mathcal{T}^{2d-2}$
with a singularity at one point, say in $\mathcal{T}^{2d-2}_{1/2}$,
and the $\downarrow$-frame is singular only
at the time reversal of that point.

The curvature term of (\ref{FKf}) splits into two terms
\begin{equation}\int_{\mathcal{T}^{2d-2}_{1/2}}\mbox{Tr}(\mathcal{F}^{d-1})
=\left[\int_{\mathcal{T}^{2d-2}_{1/2}}-\int_{\mathcal{T}^{2d-2}\backslash\mathcal{T}^{2d-2}_{1/2}}\right]
\mbox{Tr}(\mathcal{F}^{d-1}_\uparrow)
\end{equation} And the two spin components of
the Chern-Simons term $\oint_{\partial\mathcal{T}^{2d-2}_{1/2}}\mathcal{Q}_{2d-3}$
add up into \begin{equation}
2\oint_{\partial\mathcal{T}^{2d-2}_{1/2}}\mathcal{Q}^\uparrow_{2d-3}
=-2\int_{\mathcal{T}^{2d-2}\backslash\mathcal{T}^{2d-2}_{1/2}}
\mbox{Tr}(\mathcal{F}^{d-1}_\uparrow)\end{equation}
by Stokes theorem.

Combining these two, we equate (\ref{FKf}) to the spin
Chern number \begin{equation}n_\uparrow=\frac{i^{d-1}}{(d-1)!2\pi^{d-1}}
\int_{\mathcal{T}^{2d-2}}\mbox{Tr}(\mathcal{F}^{d-1}_\uparrow)\end{equation}
Time reversal requires $n_{tot}=n_\uparrow+n_\downarrow=0$ and therefore
$n_\sigma=(n_\uparrow-n_\downarrow)/2=n_\uparrow$.

\subsection{Point defects in class DIII}
\label{appendix:pdDIII}

\begin{figure}
\epsfxsize=2.5in
\epsfbox{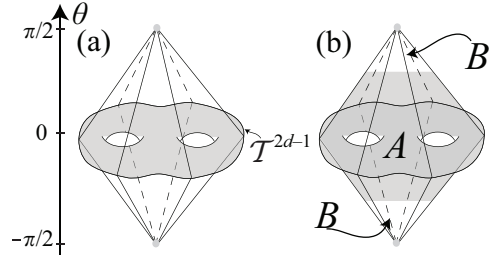}
\caption{
(a) Schematic of the suspension $\Sigma\mathcal{T}^{2d-1}$.
(b) Decomposition into patches $A$ and $B$.}
\label{susp2}
\end{figure}

The base space manifold is $\mathcal{T}^{2d-1}=T^d\times S^{d-1}$.
The Hamiltonian mapping \eqref{ctonc} relates a point defect Hamiltonian
$\mathcal{H}({\bf k},{\bf r})$ in class DIII to a line defect Hamiltonian
$\mathcal{H}({\bf k},{\bf r},\theta)=\cos\theta\mathcal{H}({\bf k},{\bf r})
+\sin\theta\Pi$ in class AII, where $\Pi=i\Theta\Xi$ is the chiral operator,
$({\bf k},{\bf r},\theta)\in\Sigma\mathcal{T}^{2d-1}$
(see fig.\ref{susp2}(a)) and $\theta$ is odd under time reversal.
The line defect Hamiltonian $\mathcal{H}({\bf k},{\bf r},\theta)$ is topologically
characterized by the generalization of (\ref{FKf}), which
was proven to be identical to the winding number (\ref{windingFK}) of the
transition function $t^{AB}$ (see fig.\ref{susp2}(b) for the definition
of patches $A$ and $B$). We will utilize this to construct a topological
invariant that characterizes point defects in class DIII.

Set $\Theta=\tau_yK$ and $\Xi=\tau_xK$ under an
appropriate choice of basis. A {\em canonical} valence frame of
$\mathcal{H}({\bf k},{\bf r},\theta)$ can be chosen to be
\begin{equation}\label{vfpDIII}u^B_+({\bf k},{\bf r},\theta)
=\left(\begin{array}{*{20}c}\sin\left(\frac{\pi}{4}-\frac{\theta}{2}\right)q({\bf k},{\bf r})\\
-\cos\left(\frac{\pi}{4}-\frac{\theta}{2}\right)\openone\end{array}\right)\end{equation}
where $q({\bf k},{\bf r})\in U(k)$ is from the canonical form of
the chiral Hamiltonian $\mathcal{H}({\bf k},{\bf r})$ in eq.\eqref{hchiral},
$\openone$ is the $k\times k$ identity matrix, and the valence frame
is non-singular everywhere except at $\theta=-\pi/2$. There is a gauge
transformation $u^B_+\to u^A=u^B_+t^{BA}$ everywhere except
$\theta=\pm\pi/2$ such that the new frame $u^A$ satisfies the gauge
condition (\ref{FKgc}).\cite{ldtopobst2} A valence frame on patch B
can be constructed by requiring $u^B_-(\theta)=\Theta u^B_+(-\theta)$
around $\theta=-\pi/2$.

The $\mathbb{Z}_2$-topology is characterized by the evenness or oddness
of the winding number of $t^{AB}$ as in \eqref{windingFK}. This can be evaluated by the integral
along the equator $\theta=0$
\begin{equation}\label{wnpDIII}\tilde\nu=\frac{(d-1)!}{(2d-1)!(2\pi i)^d}\int_{T^d\times S^{d-1}}\mbox{Tr}\left[\left(t^{AB}d(t^{AB})^\dagger\right)^{2d-1}\right]\end{equation}
where $u^A=u^B_+t^{BA}$ is a solution to the gauge condition (\ref{FKgc}),
or equivalently $t^{BA}$ satisfies
\begin{equation}\label{FKgcq}q({\bf k},{\bf r})
=t^{BA}(-{\bf k},{\bf r})\sigma_yt^{BA}({\bf k},{\bf r})^T\end{equation}
where the constant in eq.(\ref{FKgc}) is chosen to be $i\sigma_y$.

The winding number \eqref{wnpDIII} can also be expressed as a Chern-Simons
integral. \begin{equation}\tilde\nu=\frac{i^d}{d!(2\pi)^d}\int_{T^d\times S^{d-1}}
\left(\mathcal{Q}^B_{2d-1}-\mathcal{Q}^A_{2d-1}\right)\label{DIIICS1}\end{equation}
where $\mathcal{Q}^{A/B}$ are the Chern-Simons form generated by
valence frames $u^{A/B}$. Restricted to $\theta=0$, \eqref{vfpDIII} gives
$u^B({\bf k},{\bf r})=\frac{1}{\sqrt{2}}(q({\bf k},{\bf r}),-\openone)$.
Following \eqref{CS=wn/2}, the first term of \eqref{DIIICS1} equals
half of the winding number of $q$, which is guaranteed to be zero
by time reversal and particle-hole symmetries. And therefore point defects
in DIII are classified by the Chern-Simons invariant
\begin{equation}\tilde\nu=\frac{1}{d!}\left(\frac{i}{2\pi}\right)^d\int_{T^d\times S^{d-1}}
\mathcal{Q}_{2d-1}\quad\mbox{mod 2}\label{DIIICS2}\end{equation} where the
Chern-Simons form is generated by a valence frame that satisfies
the time reversal gauge constraint \eqref{FKgc}.

Note that the integrality of the Chern-Simons integral \eqref{DIIICS2} is
a result of particle-hole symmetry. Forgetting time reversal
symmetry, point defects in class D are classified by the
Chern-Simons invariant $\nu=2\tilde\nu$ \eqref{CSD} with a factor of 2.
Time reversal symmetry requires the zero modes to form Kramers doublets, and
therefore $\nu=2\tilde\nu$ must be even. A gauge transformation in
general can alter $\tilde\nu$ by any integer. Thus, similar to the
formula in class AII, the time reversal gauge constraint
\eqref{FKgc} is essential so that \eqref{DIIICS2} is non-vacuous.

\subsubsection{Fixed points formula in 1D}
\label{appendix:fixedpointformulas}

We here identify \eqref{DIIICS2}, or equivalently \eqref{wnpDIII}, to
a fixed point invariant in 1 dimension. In ref.[\onlinecite{qhz10}], Qi, Hughes and Zhang showed
that 1D TRI superconductors are $\mathbb{Z}_2$-classified by
the topological invariant
\begin{equation}\label{QHZTRIS}
(-1)^{\tilde\nu}=\frac{\mbox{Pf}(q_{k=\pi})}{\mbox{Pf}(q_{k=0})}
\exp\left(\frac{1}{2}\int_0^\pi\mbox{Tr}(q_kdq_k^\dagger)\right)\end{equation}
under the basis $\Theta=\tau_yK$ and $\Xi=\tau_xK$,
where $q_k$ is the chiral flipping operator in \eqref{hchiral}.
Time reversal and particle-hole symmetry requires $q_k=-q_{-k}^T$.
Hence the Pfaffians are well defined as $q_k$
is antisymmetric at the fixed points $k=0,\pi$.

Using the gauge condition \eqref{FKgcq}, we can expressed the
Pfaffians as $\mbox{Pf}(\Theta q_{k=0,\pi})=\det(t_{k=0,\pi})
\mbox{Pf}(\sigma_y)$, where $t_k^{BA}$ is abbreviated to $t_k$.
\begin{equation}\frac{\mbox{Pf}(q_{k=\pi})}
{\mbox{Pf}(q_{k=0})}=\exp\left(-\int_0^\pi
\mbox{Tr}\left(t_kdt_k^\dagger\right)\right)\end{equation}
Substitute \eqref{FKgcq} into the Cartan form $\mbox{Tr}(q_kdq_k^\dagger)$
gives \begin{equation}\mbox{Tr}\left(q_kdq_k^\dagger\right)
=\mbox{Tr}\left(t_{-k}dt_{-k}^\dagger\right)
+\mbox{Tr}\left(t_kdt_k^\dagger\right)\end{equation}
Combining these into \eqref{QHZTRIS},
\begin{eqnarray}(-1)^{\tilde\nu}&=&\exp\left(\frac{1}{2}\int_0^\pi
\mbox{Tr}\left(t_{-k}dt_{-k}^\dagger-t_kdt_k^\dagger\right)\right)\\
&=&\exp\left(-\frac{1}{2}\int_{-\pi}^\pi
\mbox{Tr}\left(t_kdt_k^\dagger\right)\right)\end{eqnarray}
which agrees \eqref{wnpDIII}.

\section{Invariant for fermion parity pumps}
\label{appendix:pumpinvariant}

In the appendix, we will show that a $\mathbb{Z}_2$-invariant
\eqref{FPPFK} under a particle-hole gauge constraint
topologically classified fermion parity pumps in dimension
$\delta=0$ and class D or BDI. (See sec.\ref{sec:pumpparity} for the full statement.)

\begin{figure}
\epsfxsize=2.5in
\epsfbox{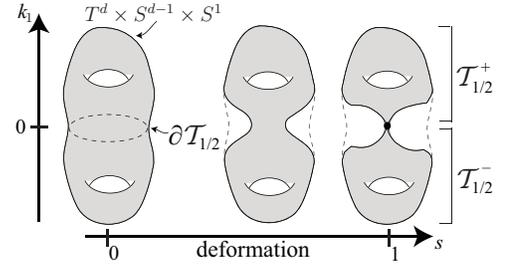}
\caption{Deformation of Hamiltonian and its base manifold so that
the boundary $\partial\mathcal{T}_{1/2}$ is shrink to a point.}
\label{defjm}
\end{figure}
We will show \eqref{FPPFK} using a construction similar to a reasoning
in Moore and Balents [\onlinecite{moorebalents07}]. We consider a deformation of the Hamiltonian
along with the base manifold, so that the boundary $\partial\mathcal{T}_{1/2}$
is deformed into a single point (see fig \ref{defjm}). Let $s\in[0,1]$
be the deformation variable, and denote $\mathcal{T}_{1/2}^+(s)$ and
$\partial\mathcal{T}_{1/2}(s)$ be the corresponding deformation slices.
Chern invariant \begin{equation}n=\frac{1}{d!}\left(\frac{i}{2\pi}\right)^d
\int_{\mathcal{T}^+_{1/2}(s=1)}\mbox{Tr}(\mathcal{F}^d)\end{equation}
integrally classifies Hamiltonians on the half manifold $\mathcal{T}^+_{1/2}(s=1)$.
Particle-hole symmetry requires opposite Chern invariant on the
other half. A different choice of deformation could only change the
Chern invariant by an even integer.\cite{FPPapp2Z} Hence the Chern
invariant modulo 2 defines a $\mathbb{Z}_2$-invariant.

The Chern integral can be further deformed and decomposed into
\begin{eqnarray}&&\int_{\mathcal{T}^+_{1/2}(s=0)}\mbox{Tr}(\mathcal{F}^d)
+\int_0^1ds\int_{\partial\mathcal{T}^+_{1/2}(s)}\mbox{Tr}(\mathcal{F}^d)
\nonumber\\&=&\int_{\mathcal{T}^+_{1/2}(s=0)}\mbox{Tr}(\mathcal{F}^d)
-\oint_{\partial\mathcal{T}^+_{1/2}(s=0)}\mathcal{Q}_{2d-1}\end{eqnarray}
where Stokes' theorem is used and the negative sign is from a
change of orientation of the boundary. This proves \eqref{FPPFK}.
The particle-hole gauge constraint is built-in since the
$G_{{\bf k},{\bf r},t}(s)$ deforms into a constant at $s=1$ while respecting
particle-hole symmetry \eqref{FPPgc} at all $s$.

\end{appendix}


\begin{thebibliography}{10}

\bibitem{tknn} D.J. Thouless, M. Kohmoto, M.P. Nightingale and M. den Nijs,
Phys. Rev. Lett. {\bf 49}, 405 (1982).

\bibitem{qizhang10} X.L. Qi and S.C. Zhang, Physics Today {\bf 63}, 33 (2010).

\bibitem{moore10} J.E. Moore, Nature {\bf 464}, 194 (2010).

\bibitem{hk10} M.Z. Hasan and C.L. Kane, arXiv:1002.3895v1 (2010).


\bibitem{km05a} C.L. Kane and E.J. Mele Phys. Rev. Lett. {\bf 95}
226801 (2005).

\bibitem{km05b}  C.L. Kane and E.J. Mele Phys. Rev. Lett. {\bf 95}
146802 (2005).


\bibitem{moorebalents07} J.E. Moore and L. Balents, Phys. Rev. B {\bf 75},
121306(R) (2007).


\bibitem{roy1} R. Roy, Phys. Rev. B {\bf 79}, 195322 (2009);
arXiv:0607531 (2006).

\bibitem{fkm07} L. Fu, C.L. Kane and E.J. Mele, Phys. Rev. Lett. {\bf
98}, 106803 (2007).

\bibitem{qihugheszhang08}
X. L. Qi, T. L. Hughes, and S. C. Zhang,Phys. Rev.
B {\bf 78}, 195424 (2008).

\bibitem{bhz06} A. Bernevig, T. Hughes and S.C. Zhang, Science {\bf 314},
1757 (2006).

\bibitem{fukane07}  L. Fu and C.L. Kane, Phys. Rev. B {\bf 76}, 045302
(2007).

\bibitem{zhang09} H. Zhang, C. X. Liu, X. L. Qi, X. Dai, Z. Fang and S. C.
Zhang, Nature Physics {\bf 5}, 438 (2009).

\bibitem{konig1} M. K\"onig, S. Wiedmann, C. Br\"une, A. Roth, H. Buhmann, L. Molenkamp, X.L.
Qi and S.C. Zhang, Science {\bf 318}, 766 (2007).

\bibitem{konig2} M. K\"onig, H. Buhmann, L. W. Molenkamp, T. Hughes, C.
X. Liu, X. L. Qi and S. C. Zhang, J. Phys. Soc. Jpn. {\bf 77}, 031007 (2008).

\bibitem{roth} A. Roth, C. Br\"une, H. Buhmann, L. W. Molenkamp, J. Maciejko,
X. L. Qi, and S. C. Zhang, Science {\bf 325}, 294 (2009).

\bibitem{hsieh08} D. Hsieh, , D. Qian, L. Wray, Y. Xia, Y. S. Hor, R. J. Cava
and M. Z. Hasan, Nature {\bf 452}, 970 (2008).

\bibitem{hsieh09a} D. Hsieh, Y. Xia, L. Wray, D. Qian, A. Pal, J. H. Dil, J.
Osterwalder, F. Meier, G. Bihlmayer, C. L. Kane, Y. S.
Hor, R. J. Cava and M. Z. Hasan, Science {\bf 323}, 919 (2009).

\bibitem{roushan09} P. Roushan, J. Seo, C. V. Parker, Y. S. Hor, D. Hsieh, D.
Qian, A. Richardella, M. Z. Hasan, R. J. Cava and A. Yazdani, Nature {\bf 460}, 1106 (2009).

\bibitem{xia09a} Y. Xia, D. Qian, D. Hsieh, L.Wray, A. Pal, H. Lin, A. Bansil,
D. Grauer, Y. S. Hor, R. J. Cava and M. Z. Hasan, Nat. Phys. {\bf 5}, 398 (2009).

\bibitem{hor09} Y. S. Hor, A. Richardella, P. Roushan, Y. Xia, J. G. Checkelsky,
A. Yazdani, M. Z. Hasan, N. P. Ong, and R. J. Cava,
Phys. Rev. B {\bf 79}, 195208 (2009).

\bibitem{chen09} Y. L. Chen, J. G. Analytis, J. H. Chu, Z. K. Liu, S. K. Mo,
X. L. Qi, H. J. Zhang, D. H. Lu, X. Dai, Z. Fang, S. C.
Zhang, I. R. Fisher, Z. Hussain, Z. X. Shen, Science {\bf 325}, 178 (2009).

\bibitem{hsieh09b} D. Hsieh, Y. Xia, D. Qian, L. Wray, J. H. Dil, F. Meier, J.
Osterwalder, L. Patthey, J. G. Checkelsky, N. P. Ong, A.
V. Fedorov, H. Lin, A. Bansil, D. Grauer, Y. S. Hor, R. J.
Cava and M. Z. Hasan, Nature {\bf 460}, 1101 (2009).

\bibitem{park10} S. R. Park, W. S. Jung, C. Kim, D. J. Song, C. Kim, S.
Kimura, K. D. Lee and N. Hur, Phys. Rev. B {\bf 81}, 041405(R) (2010).

\bibitem{alpichshev10} Z. Alpichshev, J. G. Analytis, J. H. Chu, I. R. Fisher, Y. L.
Chen, Z. X. Shen, A. Fang, and A. Kapitulnik, Phys. Rev. Lett. {\bf 104}, 016401
(2010).

\bibitem{hsieh09c} D. Hsieh, Y. Xia, D. Qian, L. Wray, F. Meier, J. H. Dil, J.
Osterwalder, L. Patthey, A. V. Fedorov, H. Lin, A. Bansil,
D. Grauer, Y. S. Hor, R. J. Cava, and M. Z. Hasan,
Phys. Rev. Lett. {\bf 103}, 146401 (2009).

\bibitem{shitade09} A. Shitade, H. Katsura, J. Kunes, X. L. Qi, S. C. Zhang and
N. Nagaosa, Phys. Rev. Lett. {\bf 102}, 256403 (2009).

\bibitem{pesin10} D. A. Pesin and L. Balents, arXiv:0907.2962.

\bibitem{chadov10} S. Chadov, X. L. Qi, J K\"ubler, G. H. Fecher, C. Felser, S. C.
Zhang, arXiv:1003.0193 (2010).

\bibitem{lin10a} H. Lin, L. A. Wray, Y. Xia, S. Jia, R. J. Cava, A. Bansil, M.
Z. Hasan, arXiv:1003.0155.

\bibitem{lin10b} H. Lin, L. A. Wray, Y. Xia, S. Y. Xu, S. Jia, R. J. Cava, A.
Bansil, M. Z. Hasan, arXiv:1003.2615.

\bibitem{lin10c}
H. Lin, L. A. Wray, Y. Xia, S. Y. Xu, S. Jia, R. J. Cava, A.
Bansil, M. Z. Hasan, arXiv:1004.0999.

\bibitem{yan10}
B. Yan, C. X. Liu, H. J. Zhang, C. Y. Yam, X. L. Qi, T.
Frauenheim, S. C. Zhang, arXiv:1003.0074.

\bibitem{roy08} R. Roy, arXiv:0803.2868 (unpublished).

\bibitem{schnyder08} A. P. Schnyder, S. Ryu, A. Furusaki and A. W. W. Ludwig,
Phys. Rev. B {\bf 78}, 195125 (2008);
A.P. Schnyder, S. Ryu, A. Furusaki and A. W. W. Ludwig,
AIP Conf. Proc. {\bf 1134}, 10 (2009).

\bibitem{kitaev09} A. Kitaev, AIP Conf. Proc. {\bf 1134}, 22 (2009).

\bibitem{qi09} X.L. Qi, T. L. Hughes, S. Raghu and S. C. Zhang,
Phys. Rev. Lett. {\bf 102}, 187001 (2009).

\bibitem{kitaev00} A. Kitaev, arXiv:cond-mat/0010440 (unpublished).

\bibitem{readgreen} N. Read, and D. Green, Phys. Rev. B {\bf 61}, 10267 (2000).

\bibitem{mackenzie03} A. P. Mackenzie and Y. Maeno, Rev. Mod. Phys. {\bf 75},
657 (2003).

\bibitem{grinevich88} P. G. Grinevich and G. E. Volovik,
J. Low Temp. Phys. {\bf 72}, 371 (1988).

\bibitem{volovik03} G. E. Volovik, {\it The Universe in a Helium Droplet},
(Clarendon, Oxford, 2003).

\bibitem{silaev10} M. A. Silaev, G. E. Volovik, arXiv:1005.4672.

\bibitem{volovik09} G. E. Volovik, JETP Lett. {\bf 90}, 587 (2009).

\bibitem{altland97}
A. Altland and M. R. Zirnbauer, Phys. Rev. B {\bf 55}, 1142 (1997).

\bibitem{horava} P. Horava, Phys. Rev. Lett. {\bf 95}, 016405 (2005).

\bibitem{jackiwrebbi}
R. Jackiw and C. Rebbi,Phys. Rev. D {\bf 13}, 3398 (1976).

\bibitem{jackiwrossi}
R. Jackiw and P. Rossi, Nucl. Phys. B, {\bf 190} 681 (1981).

\bibitem{ssh}
W. P. Su, J. R. Schrieffer, and A. J. Heeger,
Phys. Rev. Lett. {\bf 42}, 1698 (1979).

\bibitem{fukane08}
L. Fu and C. L. Kane, Phys. Rev. Lett. {\bf 100}, 096407 (2008).

\bibitem{sau10} J. D. Sau, R. M. Lutchyn, S. Tewari, S. Das Sarma,
Phys. Rev. Lett. {\bf 104}, 040502 (2010).

\bibitem{alicea10} J. Alicea,  Phys. Rev. B {\bf 81}, 125318 (2010).

\bibitem{lutchyn10} R. M. Lutchyn, J. D. Sau, S. Das Sarma, arXiv:1002.4033
(2010).

\bibitem{oreg10} Y. Oreg, G. Refael and F. von Oppen,
arXiv:1003.1145 (2010).

\bibitem{teokane10} J. C. Y. Teo and C. L. Kane,
Phys. Rev. Lett. {\bf 104}, 046401 (2010).

\bibitem{nakahara} M. Nakahara, ``Geometry, Topology and Physics,"
Adam Hilger (1990).

\bibitem{goldstone} J. Goldstone and F. Wilczek, Phys. Rev. Lett.
{\bf 47}, 986 (1981).

\bibitem{weinberg} E. J. Weinberg, Phys. Rev. D {\bf 24}, 2669 (1981).

\bibitem{witten} E. Witten, Phys. Lett. {\bf 117B}, 324 (1982).

\bibitem{witten82} E. Witten, J. Diff. Geom. {\bf 17}, 611 (1982).

\bibitem{davis} S.C. Davis, A.C. Davis, W.B. Perkins, Phys. Lett. {\bf 408B}, 81
(1997).



\bibitem{thouless} D. J. Thouless, Phys. Rev. B {\bf 27}, 6083
(1983).

\bibitem{thoulessniu} Q. Niu and D.J. Thouless, J. Phys. A {\bf 17}
2453 (1984).

\bibitem{freedman10}
M. Freedman, M. B. Hastings, C. Nayak, X. L. Qi, K. Walker and Z. Wang,
arXiv:1005.0583 (2010).

\bibitem{karoubi} M. Karoubi, {\it K-theory: an introduction}, Springer-Verlag,
1978.

\bibitem{lawson} H. B. Lawson, M. L. Michelsohn, {\it Spin Geometry},
Princeton University Press (1989).

\bibitem{atiyah94} M. Atiyah, {\it K Theory}, Westview Press (1994).

\bibitem{bott} R. Bott, Ann. Math. {\bf 70}, 313 (1959).

\bibitem{JMilnor}  J. Milnor, {\it Morse Theory},
Princeton University Press, (1963).

\bibitem{blount} E. I. Blount, Solid State Phys. {\bf 13}, 305 (1962).

\bibitem{ryu10}
S. Ryu, A. Schnyder, A. Furusaki, A. Ludwig, arXiv:0912.2157 (2009).

\bibitem{li10}
R. Li,  J. Wang, X. L. Qi and S. C. Zhang,
Nature Physics {\bf 6}, 284 (2010).

\bibitem{mong10}
R. S. K. Mong, A. M. Essin, J. E. Moore
Phys. Rev. B {\bf 81}, 245209 (2010).

\bibitem{volovik10}
G. E. Volovik, Pis'ma ZhETF {\bf 91} 61 (2010); arXiv:1001.1514.

\bibitem{fukane06}
L. Fu and C. L. Kane, Phys. Rev. B {\bf 74}, 195312 (2006).

\bibitem{ran09}
Y. Ran, Y. Zhang and A. Vishwanath, Nat. Phys. {\bf 5}, 298 (2009).

\bibitem{laughlin} R. B. Laughlin, Phys. Rev. Lett. {\bf 80}, 5188 (1998).

\bibitem{thankroman} We
thank Professor R. Jackiw for asking the question that inspired this calculation.

\bibitem{pontrjagin} L.S. Pontrjagin, Rec. Math. [Mat. Sbornik] N.S. {\bf 9},
331 (1941); http://mi.mathnet.ru/eng/msb6073.

\bibitem{jaykka} J. J\"aykk\"a and J. Hietarinta, Phys. Rev. D {\bf 79}, 125027
(2009).

\bibitem{kapitanski} L. Kapitanski, in London Mathematical Society Durham
Symposium,
(unpublished),
http://www.
maths.dur.ac.uk/events/Meetings/LMS/200/OTSA/.

\bibitem{qhz10} X. L. Qi, T. L. Hughes and S. C. Zhang,
Phys. Rev. B {\bf 81}, 134508 (2010).

\bibitem{fukane09a} L. Fu, and C. L. Kane, Phys. Rev. B {\bf 79}, 161408(R)(2009).

\bibitem{fukui09} T. Fukui and T. Fujiwara, J. Phys. A {\bf 42}, 362003 (2009).

\bibitem{ran10b} Y. Ran, P. Hosur, A. Vishwanath, arXiv:1003.1964 (2010).

\bibitem{laughlin81}  R. B. Laughlin, Phys. Rev. B {\bf 23}, 5632(R)
(1981).

\bibitem{fukui10a}  T. Fukui and T. Fujiwara, J. Phys. Soc. Jpn. {\bf 79},
033701 (2010).

\bibitem{fukui10b}  T. Fukui, arXiv:1003.4814v1 (2010).

\bibitem{ldtopobst} Bulk topology is a lower dimenional topological obstruction
coming purely from the momentum part $T^d$ of the base space $T^d\times S^{d-2}$.
One could mathematically remove the bulk topological obstruction by adding a
defectless Hamiltonian $\widetilde{\mathcal{H}}({\bf k},{\bf r})
=\mathcal{H}({\bf k},{\bf r})\oplus\mathcal{H}_0({\bf k})$.

\bibitem{ldtopobst2} Again, assume the bulk has trivial topology.
Or otherwise remove the topological obstruction mathematically by adding
a defectless Hamiltonian.

\bibitem{FPPapp2Z} Deformation of Hamiltonians on $\partial\mathcal{T}_{1/2}(s)$, for $s\in[0,1]$
has dimension $\delta=(d-1)-(d-1+1+1)=-2$, and is classified by $2\mathbb{Z}$.


\end{thebibliography}
\end{document}